\documentclass[pra,aps,10pt,superscriptaddress,twocolumn,floatfix,showpacs]{revtex4-1}
\usepackage{amsmath,amsfonts,dsfont,graphicx,amssymb,microtype,braket,xcolor,upgreek,mathtools,physics}
\usepackage{placeins}
\usepackage[colorlinks, linkcolor=blue, citecolor=blue, urlcolor=blue, breaklinks=true]{hyperref}
\DeclareGraphicsExtensions{.pdf}

\usepackage[toc,page]{appendix}
\newcommand{\td}{\mathrm{d}}

\newcommand{\ts}[1]{_\text{#1}}

\newcommand{\s}{\sigma}

\newcommand{\me}{\mathrm{e}}
\newcommand{\mi}{\mathrm{i}}
\usepackage{physics}
\newcommand{\sumbyparity}[3]{\sideset{^{\mathrm{p}}}{}{\sum}_{#1=#2}^{#3}}
\newcommand{\sumbyparitysymb}[1]{\sideset{^{\mathrm{p}}}{}{\sum}^{#1}}

\begin{document}

\title{Multidimensional semiclassical single- and double-quantum spectroscopy of anharmonic molecular polaritons}
\author{Michael Reitz}
\email{mireitz@ucsd.edu}
\affiliation{Department of Chemistry and Biochemistry, University of California San Diego, La Jolla, California 92093, USA}
\author{Harsh Bhakta}
\affiliation{Department of Chemistry and Biochemistry, University of California San Diego, La Jolla, California 92093, USA}
\author{Wei Xiong}
\affiliation{Department of Chemistry and Biochemistry, University of California San Diego, La Jolla, California 92093, USA}
\affiliation{Program in Materials Science and Engineering, University of California San Diego, La Jolla, California 92093, USA}

\author{Joel Yuen-Zhou}
\email{joelyuen@ucsd.edu}
\affiliation{Department of Chemistry and Biochemistry, University of California San Diego, La Jolla, California 92093, USA}

\date{\today}

\begin{abstract}
 We present a general and efficient approach to compute phase-resolved multidimensional spectra of anharmonic molecular polaritons, based on a semiclassical evolution of the molecular Hamiltonian and cavity field in the large-$\mathcal{N}$ limit of many molecules coupled to a confined photonic mode. By systematically expanding the response in both amplitudes and phases of the input fields, our method enables a transparent and computationally simple construction of phase-cycled two-dimensional single- and double-quantum polariton spectra from the underlying nonlinear signal components. Here, phase cycling acts as an analogue of phase matching with oblique pulses, allowing for the isolation of the contributing nonlinear pathways in Liouville space. We specialize to vibrational polaritons and benchmark the method through direct comparison with experimentally measured single-quantum spectra, providing an explanation for the longstanding puzzle of the polariton bleach effect observed at short waiting times. Further, we show how the imprint of various types of anharmonicities on the double-excitation manifold can be directly probed and analyzed through double-quantum coherence spectroscopy. Taken together, our results establish a practical and powerful framework for the modeling and interpretation of nonlinear spectroscopic experiments on strongly coupled light-matter platforms and for guiding the design of cavity-enhanced molecular platforms.
\end{abstract}

\maketitle


\section{Introduction}

Multidimensional spectroscopy is a powerful and well-established method for investigating anharmonicities, interstate couplings and energy transfer processes, among other properties, in complex molecular and solid-state systems, thereby providing information that is inaccessible to linear spectroscopies \cite{mukamel2000multidimensional, khalil2001signatures, jonas2003two, read2007cross, hochstrasser2007two, kim2009two, maiuri2020ultrafast}. The basic idea is to interrogate a system with a sequence of ultrafast laser pulses to excite molecular levels to higher manifolds or prepare various coherence and population states, and then to extract and analyze the resulting nonlinear signal as a function of the interpulse delays. To this end, a wide variety of experimental setups, phase-matching geometries, and phase cycling schemes have been developed \cite{Hamm_Zanni_2011, yuenzhou2014ultrafast}. Originally developed in the context of nuclear magnetic resonance to investigate spin dynamics \cite{ernst1990principles}, these concepts were later transferred to infrared (vibrational) as well as electronic spectroscopy \cite{tanimura1993two, hybl1998two, jonas2003optical}. Advances in laser technology have pushed the achievable time and frequency resolutions, thereby enabling direct access to ultrafast molecular and electronic dynamics, such as those governing photosynthetic light-harvesting \cite{brixner2005two}, chemical reactions \cite{attar2017femtosecond}, water and other liquid dynamics \cite{fecko2003ultrafast, zheng2007ultrafast}, protein conformations \cite{Hamm1998structure, chung2007transient}, or superconductivity \cite{Wu2024ultrafast, salvador2024principles}. 

Yet, the above-mentioned multidimensional spectroscopy studies were applied and limited to weakly coupled systems. More recently, the techniques of ultrafast and multidimensional spectroscopy have been applied to strongly coupled light-matter systems, in which resonant interactions between confined electromagnetic modes and vibrational or electronic excitations (e.g., of molecules) lead to the formation of hybrid light-matter states known as polaritons \cite{weisbuch1992observation, Byrnes2014exciton, xiong2023molecular, yuenzhou2019polariton, Xiang2024molecular}.  The first
experiments on ultrafast polariton dynamics focused on characterizing the coherent Rabi oscillations and polariton relaxation dynamics in a variety of cavity and plasmonic
settings \cite{virgili2011ultrafast, vasa2013realtime, balci2014probing, takemura2015dephasing, dunkelberger2016modified, FinkelsteinShapiro2021understanding, kuttruff2023sub}. Two-dimensional (2D) spectroscopy of molecular ensembles strongly coupled to infrared or optical cavities has provided additional insight by offering a state-resolved view of polariton dynamics \cite{takemura2015two, xiang2018twodimensional,xiang2019manipulating, timmer2023plasmon, russo2024direct,Chen2025tracking, sufrin2026phase}, for instance revealing signatures of intermediate states that mediate polariton relaxation \cite{xiang2019state, Hirschmann2025role}. Other experimental works have explored polariton-mediated energy transfer \cite{xiang2020intermolecular, mewes2020energy, chen2022cavity}, the role of anharmonicities \cite{sufrin2024probing} or many-particle correlations \cite{quiros2026resolving, wen2013influence}, and have exploited phase-matching conditions to disentangle the polariton response \cite{michail2024addressing}. Both experimental and theoretical studies have aimed to interpret the observed signals in terms of the nonlinear ladder of polaritonic excitations \cite{delpo2020polariton,fassioli2021femtosecond,autry2020excitation,Grafton2021excited,fumero2025biexciton,YuenZhou2025PolaritonChemistry}. In parallel, theoretical efforts have been devoted to the development of analytical and numerical frameworks for nonlinear and multidimensional cavity spectroscopy \cite{ribeiro2018theory,mondal2023quantum,zhang2023multidimensional,shah2023qudpy,gallego2024coherent,schnappinger2024disentangling,philipp2025line,dewit2025process, mondal2025polariton}, many of which are however restricted to systems containing a relatively small number of molecules. Despite these efforts, a consistent interpretation of the dynamical nonlinear polariton response—especially at short pulse delay times, i.e., before dephasing occurs—as well as the unambiguous identification of genuinely polaritonic (as opposed to reservoir) features, remain elusive and subjects of active debate~\cite{renken2021untargeted,duan2021isolating,simpkins2023comment,Duan2023reply,pyles2024revisiting}.\\

\begin{figure}[b]
    \centering
    \includegraphics[width=0.85\columnwidth]{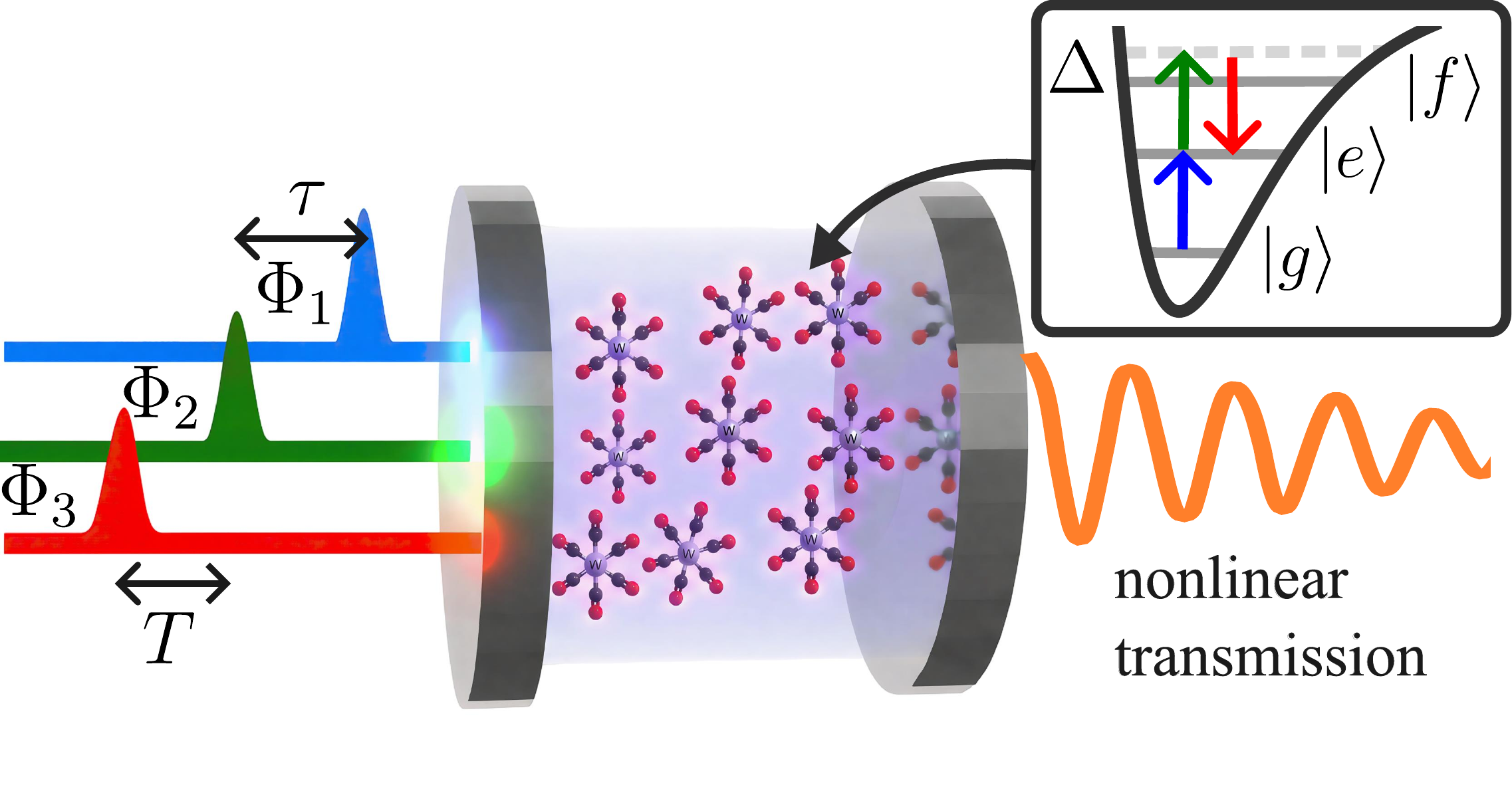}
    \caption{Schematics. An infrared optical cavity containing an ensemble of molecules (illustrated by W(CO)$_6$ as, e.g., used in Refs.~\cite{xiang2018twodimensional, xiang2019manipulating}) is driven by a train of three input pulses (two pump, one probe) tagged by phases $\Phi_j$ and separated by time intervals $\tau$ and $T$. The cavity selectively couples to a molecular vibrational mode, which is modeled as a three-level system (3LS) with anharmonic shift $\Delta$ of the doubly excited state (see inset, which depicts a 2QC process). The nonlinear cavity transmission as a function of the pulse delay times is used to compute the 2D spectra.}
    \label{schem}
\end{figure}

We have previously introduced a general and versatile method for nonlinear polariton spectroscopy in Ref.~\cite{reitz2025nonlinear}. The formalism is based on a semiclassical evolution of the coupled light-matter system and is related to Maxwell-Schrödinger (Maxwell-Liouville) approaches that are well established in semiconductor physics, nanoplasmonics, and related fields \cite{jahnke1996excitonic,lopata2009multiscale,sukharev2011numerical,Sukharev2017Exciton, jirauschek2019optoelectronic, bonafe2025full}. In addition, the method makes use of a perturbative expansion of both the electromagnetic field and the material degrees of freedom in the input pulse amplitudes, combined with a discrete Fourier transform in the pulse phases (analogous to phase matching with oblique pulses), so that the resulting phase-tagged equations of motion (EoM) can be systematically propagated up to a certain order in the nonlinear response. Importantly, this framework enables us to describe the time-dependent nonlinear polariton response across all relevant timescales, extending beyond the stationary regime dominated by dephased dark exciton states which has been previously understood theoretically~\cite{ribeiro2018theory, ribeiro2021enhanced}. We have also recently extended this framework to multimode cavities, enabling the study of polariton transport and spatially-resolved nonlinear phenomena inside cavities~\cite{fowlerwright2025mapping}. In the present work, we further extend this approach to multidimensional nonlinear spectroscopy, therefore enabling the systematic computation of phase-cycled 2D cavity spectra while remaining computationally efficient in the large-$\mathcal{N}$ limit of many molecules.\\

This paper is structured as follows. In Sec.~\ref{sec:method}, we present the theoretical framework underlying our approach, which is based on a semiclassical light–matter evolution combined with a perturbative expansion in the pulse amplitudes and phases, and outline its application to the computation of phase-cycled 2D spectra. In Sec.~\ref{sec:results}, we then compare the method against experimental measurements for single-quantum (1Q) spectra of vibrational polaritons at short and long waiting times (as compared to the polariton lifetime) and show how our method can explain the so-called polariton bleach effect observed at short waiting times, via the inclusion of excitation-induced dephasing (EID)~\cite{xiang2019manipulating}. Finally, we move to double-quantum coherence (2QC) spectroscopy and discuss the imprint of mechanical and electrical anharmonicities, i.e., anharmonicities in the dipole moment.

\section{Method}
\label{sec:method}

\begin{figure*}[t]
    \centering
    \includegraphics[width=1.0\textwidth]{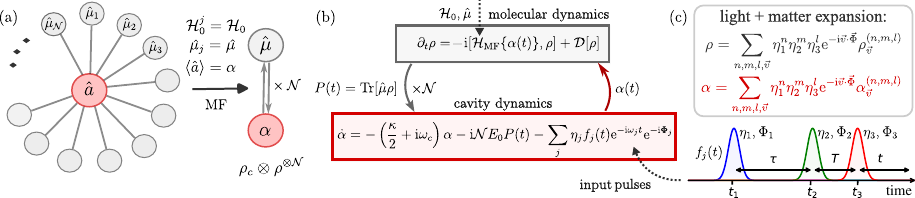}
    \caption{Overview of the semiclassical perturbative method for multidimensional cavity spectroscopy. (a) Factorizability of light and matter. The many-body system of $\mathcal{N}$ molecules described by dipole operators $\hat{\mu}_j$ coupled to a cavity mode $\hat{a}$ is reduced to a much simplified mean-field system in the large-$\mathcal{N}$ limit under the assumption $\mathcal{H}_0^j=\mathcal{H}_0$, $\hat{\mu}_j=\hat{\mu}$. (b) Schematic illustration of the self-consistent, semiclassical method. The (single-molecule) matter component is evolved quantum mechanically, from which the molecular polarization feeding into the cavity field $\alpha (t)$ is computed. In turn, the classical cavity field feeds back into the evolution of the molecular density matrix. (c) Sketch of pulse arrival times and time intervals (bottom). The time intervals are commonly referred to as excitation ($\tau$), waiting (or double-coherence) ($T$), and detection ($t$) times. Finally, the cavity field and density matrix are expanded in terms of the input pulse amplitudes and phases $\eta_j$, $\Phi_j$ (top), yielding a closed set of equations up to a certain nonlinear order. }
    \label{fig1}
\end{figure*}

This section introduces our method for  2D cavity spectroscopy in a stepwise manner. The overall basic workflow that underlies the computation of the 2D spectra is illustrated in Figs.~\ref{fig1}(a)-(c) as well as Fig.~\ref{fig2}. While we will specialize to vibrational polaritons in infrared cavities, let us emphasize here that the method is fully general and may be equally applied to systems under electronic strong coupling as well. For instance, the anharmonic three-level system (3LS) considered here can also serve as an effective model for biexcitons in semiconductors, where the biexciton binding energy leads to an anharmonic shift of the doubly excited state~\cite{borri2000biexcitons, stone2009two}. We start by introducing the basic model of $\mathcal{N}$ molecules coupled to the infrared photonic mode of a cavity which is driven by a set of input pulses in Sec.~\ref{sec:model}. Under the assumption of factorizability between light and matter in the large-$\mathcal{N}$ limit, we then move to the semiclassical description in Sec.~\ref{sec:semi}, enabling an efficient computation of the total nonlinear response which does not scale with the number of molecules $\mathcal{N}$. We then perform  a perturbative expansion in the input pulse amplitudes combined with a Fourier expansion in the pulse phases in Sec.~\ref{sec:expansion} which enables a systematic analysis of the nonlinearities underlying the 2D spectra as well as derivation of the contributing double-sided Feynman diagrams. The evolution in Liouville space is discussed in Sec.~\ref{sec:Liouville}. Finally, we discuss how one can obtain the 2D differential transmission (DT) spectra of the cavity from the perturbative equations in Sec.~\ref{sec:2dspectra}.

\subsection{Model}
\label{sec:model}

We consider a (single-mode) optical cavity at frequency $\omega_c$ containing an ensemble of $\mathcal{N}$ molecules which is driven by a train of three input pulses (see sketch in Fig.~\ref{schem}), a minimum requirement for performing 2D spectroscopy.

We start by describing the coupled cavity-molecule system. Each molecule $j$ is described by a Hamiltonian $\mathcal{H}_0^j$ and a corresponding dipole operator $\hat{\mu}_j=\hat{\mu}_j^{(+)}+\hat{\mu}_j^{(-)}$. Here, $\hat{\mu}_j^{(+)}$ and $\hat{\mu}_j^{(-)}$ denote the positive- and negative-frequency components of the dipole operator, corresponding, respectively, to molecular excitation and de-excitation processes. The Hamiltonian describing the interaction of the full coupled light-matter system is given in the rotating-wave approximation (RWA) by (setting $\hbar = 1$ here and in the following)
\begin{align}
\mathcal{H}
=\omega_c\hat{a}^\dagger\hat{a}
+\sum_{j=1}^\mathcal{N}\mathcal{H}_0^j
+E_0\sum_{j=1}^\mathcal{N}\left(\hat{a}\hat{\mu}_j^{(+)}+\hat{a}^\dagger\hat{\mu}_j^{(-)}\right),
\end{align}
where $\hat{a}$ and $\hat{a}^\dagger$ are the annihilation and creation operators of the cavity photon mode, respectively, and $E_0=\sqrt{\omega_c/(2\epsilon_0\mathcal{V})}$ sets the coupling strength to the cavity mode, with $\mathcal{V}$ the mode volume and $\epsilon_0$ the vacuum permittivity. We assume that the system is not in the ultrastrong-coupling regime, such that the RWA and standard input-output theory remain valid \cite{steck2007quantum, ciuti2006input}.

To illustrate the method, throughout the article, we consider an ensemble of (identical) 3LSs, each described by
\begin{align}
\mathcal{H}_0^j=\omega_g \ket{g}_j\bra{g}_j+\omega_e\ket{e}_j\bra{e}_j+\omega_f\ket{f}_j\bra{f}_j,
\end{align}
and characterized by a dipole operator
\begin{align}
\hat\mu_j=\mu_{ge}^{\phantom{j}}\hat{\sigma}_{ge}^j+\mu_{ef}^{\phantom{j}}\hat{\sigma}_{ef}^j+\mathrm{H.c.},
\end{align}
where $\hat{\sigma}_{ge}^j=\ket{g}_j\bra{e}_j$ and $\hat{\sigma}_{ef}^j=\ket{e}_j\bra{f}_j$ are the corresponding lowering operators for the $j\mathrm{th}$ molecule. Only the $g\!\leftrightarrow\! e$ and $e\!\leftrightarrow\! f$ transitions are dipole-allowed, reflecting the ladder structure of a single vibrational mode. Such a 3LS provides an effective model for a vibrational mode with anharmonicity (e.g., a Morse oscillator), where the states $\ket{g}_j, \ket{e}_j, \ket{f}_j$ correspond to the vibrational ground, first excited, and second excited states.  Since we focus on calculating third-order response functions in the following, truncating the vibrational Hilbert space at the two-excitation level is sufficient for capturing all relevant nonlinear pathways.  Alternatively, the same physics may be described in terms of a bosonic vibrational mode with a Kerr-type nonlinearity, whose lowest three Fock states map directly onto the Hamiltonian $\mathcal{H}_0^j$ \cite{ribeiro2018theory}. The harmonic limit is recovered when $(\omega_f-\omega_e)=(\omega_e-\omega_g)$ 
and the dipole moments satisfy $\mu_{ef}=\sqrt{2}\,\mu_{ge}$. 
Deviations from this ideal harmonic behavior quantify the anharmonicity of the vibrational mode. 
In particular, \textit{mechanical} anharmonicity is characterized by the frequency shift 
$\Delta = (\omega_f-\omega_e) - (\omega_e-\omega_g)$, 
which measures the departure from equal level spacing. 
In addition, \textit{electrical} anharmonicity describes deviations of the transition dipole moments 
from the harmonic scaling, which can be parameterized as 
$\mu_{ef} = \sqrt{2}\,\mu_{ge}(1+\delta)$, 
where $\delta$ quantifies the relative correction to the harmonic dipole ratio \cite{herzberg1939molecular}. 

Next, we consider the interaction of the cavity mode with the external driving field produced by the three external laser pulses. The pulses enter the cavity mode by transmitting through the transmission window of the cavity resonances (cavity linewidth $\kappa$). In the RWA, the excitation of the cavity mode is described by
\begin{align}
\mathcal{H}\ts{drive}
=\mi\sum_{j=1}^{3}\eta_j f_j(t-t_j)
\left(
\hat{a}^\dagger\,\me^{-\mi\omega_{\ell,j} t}\,\me^{-\mi \Phi_j}
- \mathrm{H.c.}
\right),
\end{align}
where $\eta_j$, $f_j$, $t_j$, $\omega_{\ell, j}$, $\Phi_j$ describe the amplitude, envelope, arrival time, carrier frequency and phase of the $j$th pulse, respectively. Note that, in contrast to the usual dipole interaction $-\hat{\mu}\cdot E(t)$ commonly used in nonlinear spectroscopy \cite{mukamel1995principles}, the external field here couples to the cavity photon mode via the creation and annihilation operators $\hat a^\dagger$ and $\hat a$ and the molecules are only driven indirectly via the cavity mode.  We note that alternatively, one could also excite the molecules directly through the side of the cavity \cite{zhang2023multidimensional}, which is equivalent to exciting the system with high in-plane momentum. Likewise, one can directly excite and probe molecular transitions through spectral transmission windows of the cavity, where the optical field is not strongly cavity-confined~\cite{McKillop2026direct}. However, we focus here on driving the cavity field directly, as this is the configuration most commonly used in experiments. For the presented results, we will assume (identical) Gaussian envelopes for the pulses $f_j(t)=f(t) = (2\pi \tau_w^{2})^{-1/2}\exp[-t^{2}/(2\tau_w^{2})]$ with identical pulse widths $\tau_w$ and carrier frequencies $\omega_{\ell,j}=\omega_\ell$. The time delays between pulses define the conventional 2D spectroscopy time variables: the excitation time $\tau = t_2 - t_1$ and the waiting (or double-coherence) time $T = t_3 - t_2$ [see Fig.~\ref{schem}]. In the limiting case $\tau = 0$, the first two pulses become temporally coincident, thereby reducing to the standard two-pulse pump-probe spectroscopy. Importantly, in addition, we assign an additional global phase to each pulse, $\Phi_j$, which will allow us to keep track of distinct Liouville-space pathways via phase cycling and thereby isolate different contributions to the nonlinear response \cite{yuenzhou2014ultrafast, tian2003femtosecond, tan2008theory}. Experimentally, such controlled phase shifts are routinely implemented using acousto-optic modulators (AOMs) or spatial light modulators (SLMs), which impose a well-defined radio-frequency shift on each pulse \cite{tekavec2007fluorescence}.

The full dynamics of the coupled cavity-molecule system is then described by a master equation for the total density operator $\rho_\mathrm{tot}$ of the entire light-matter system,
\begin{align}
\partial_t \rho_\mathrm{tot}
= -\mi\bigl[\mathcal{H}+\mathcal{H}\ts{drive},\rho_\mathrm{tot}\bigr]
+\mathcal{D}[\rho_\mathrm{tot}],
\end{align}
where the dissipator $\mathcal{D}[\cdot]$ accounts for irreversible loss processes. We explicitly consider cavity photon decay at rate $\kappa$,
\begin{align}
\mathcal{D}[\rho_\mathrm{tot}]
= \kappa\left(
\hat{a}\rho_\mathrm{tot}\hat{a}^\dagger-\tfrac{1}{2}\{\hat{a}^\dagger\hat{a},\rho_\mathrm{tot}\}
\right),
\end{align}
and treat homogeneous linewidth broadening via pure dephasing. This is detailed in the Supplementary Information (SI)~\ref{sec:dephasing} and is the dominant incoherent process for systems such as W(CO)$_6$ in hexane considered below. Other dissipative molecular mechanisms, such as intermolecular vibrational redistribution (IVR) may also be straightforwardly incorporated if required.

\subsection{Semiclassical light-matter evolution}
\label{sec:semi}

The main assumption of the mean-field approach is that, in the large-$\mathcal{N}$ limit, the total light-matter density matrix factorizes into a product state between the cavity field and the molecular ensemble \cite{fowler2022efficient, Mori2013, carollo2021exactness},
\begin{align}
\rho_\mathrm{tot}=\rho_c\otimes \bigotimes_{j=1}^{\mathcal{N}}\rho_{m}^j.
\end{align}
Since all molecules are taken to be identical and experience the same mean cavity field, we can set 
$\rho_{m}^j\equiv \rho$ for all $j$, i.e.,
\begin{align}
\rho_\mathrm{tot}=\rho_c\otimes \rho^{\otimes \mathcal{N}}.
\end{align}
This can also be straightforwardly generalized to disordered, i.e., inhomogeneously broadened molecular ensembles (by grouping the ensemble into subensembles) \cite{reitz2025nonlinear} or to multimode cavities \cite{fowlerwright2025mapping}; however, for clarity of presentation, we restrict ourselves here to identical molecules in a single-mode cavity, and neglect spatial variations of the cavity field that would give rise to position-dependent couplings. This factorization allows the cavity field and the molecular ensemble to be treated as two coupled but separately evolving subsystems.

Within this mean-field picture, the matter part evolves under an effective single-molecule Hamiltonian [see sketch in Fig.~\ref{fig1}(a)]
\begin{align}
\mathcal{H}\ts{MF}=\mathcal{H}_0 + \mathcal{H}\ts{int},
\end{align}
where $\mathcal{H}_0=\omega_e\ket{e}\bra{e}+\omega_f\ket{f}\bra{f}$ is the bare three-level molecular Hamiltonian introduced in Sec.~\ref{sec:model}  (we set $\omega_g=0$ here and in the following without loss of generality), and $\mathcal{H}\ts{int}=E_0\bigl[\alpha(t)\hat{\mu}^{(+)}+\alpha^*(t)\hat{\mu}^{(-)}\bigr]$
describes the interaction with the cavity mode through the collective mean-field cavity amplitude $\expval{\hat{a}}=\alpha (t)$. The mean-field treatment is  equivalent to assuming that the cavity mode remains in a coherent state at all times, such that its quantum fluctuations can be neglected.

The mean-field Hamiltonian governs the evolution of the (now single-particle) molecular density matrix as
\begin{align}
\label{eq:meanfield}
\dot{\rho}(t)=-\mi\bigl[\mathcal{H}\ts{MF},\rho(t)\bigr] + \mathcal{D}[\rho(t)],
\end{align}
where the dissipator $\mathcal{D}[\rho]$ describes the molecular dephasing. Note that no explicit driving term appears here, since the molecules are only driven indirectly through the cavity field. In the following, we map this equation of motion to Liouville space, which converts the master equation into a linear differential equation for a vectorized density matrix. This so-called vectorization \cite{mukamel1995principles, amshallem2015approaches} greatly simplifies the perturbative treatment in the following. The vectorized equation of motion can be expressed as
\begin{align}
\label{eq:meanfieldliouville}
\dot{{\vec{\rho}}}(t)=-\mi\left(\mathcal{L}_0+\mathcal{L}\ts{int}(t)+\mathcal{L}_{\mathcal{D}}\right)\vec{\rho}(t),
\end{align}
where the Liouvillians driving the molecular evolution are defined by the mapping
\begin{subequations}
\label{eq:liouvillians}
\begin{align}
[\mathcal{H}_0,\rho] 
&\;\longrightarrow\; 
\mathcal{L}_0\,\vec{\rho}, \\
[\mathcal{H}\ts{int}(t),\rho] 
&\;\longrightarrow\; 
\mathcal{L}\ts{int}(t)\,\vec{\rho}, \\
\mathcal{D}[\rho] 
&\;\longrightarrow\; 
-\mi \mathcal{L}_{\mathcal{D}}\,\vec{\rho}.
\end{align}
\end{subequations}
Explicitly, the interaction Liouvillian is given by
\begin{equation}
\mathcal{L}_{\mathrm{int}}(t)
=
E_0\!\left[
\alpha(t)\,\mathcal{L}_{\mu^{(+)}}
+
\alpha^*(t)\,\mathcal{L}_{\mu^{(-)}}
\right],
\end{equation}
where we define $[\hat{\mu}^{(\pm)},\rho]\to \mathcal{L}_{\mu^{(\pm)}}\,\vec{\rho}$.  In Liouville space, the commutator structure
accounts for interactions acting on either the ket or bra side of the density
matrix.

Considering the decay of the cavity mode, the classical equation of motion for the cavity field amplitude is given by:
\begin{equation}
\label{eq:cavitymean}
\begin{split}
\dot{\alpha}(t) ={}& 
-\!\left(\frac{\kappa}{2} + \mi\omega_{c}\right)\alpha(t)
 - \mi\,\mathcal{N} E_{0}\, P(t) \\
&\quad - \sum_{j=1}^{3} 
\eta_{j}\, f_j(t - t_{j})\, 
\me^{-\mi \omega_{\ell,j} t}\,
\me^{-\mi\Phi_{j}},
\end{split}
\end{equation}
where the molecular polarization $P(t)=\mathrm{Tr}\!\left[\hat{\mu}\,\rho(t)\right]=P_{ge}(t)+P_{ef}(t)$, consisting of the contributions of both transitions,
provides the nonlinear feedback into the cavity field that mediates the collective vibropolaritonic response. Together with the molecular master equation \eqref{eq:meanfield}, this establishes a self-consistent light-matter evolution in which the cavity field influences the molecules and is simultaneously shaped by the induced polarization acting back onto the cavity field [see Fig.~\ref{fig1}(b)]. This set of equations forms the basis for the perturbative expansion outlined in the following section.

\subsection{Phase-resolved perturbative expansion of light-matter dynamics}
\label{sec:expansion}

While the coupled mean-field Eqs.~\eqref{eq:meanfield} and \eqref{eq:cavitymean} allow for the efficient numerical simulation of the many-body system, they do not yet provide any direct physical insight into the structure of the underlying nonlinear optical processes. To reveal this structure, it is useful, following the standard approach in nonlinear spectroscopy \cite{mukamel1995principles}, to perform a perturbative expansion of all dynamical quantities in the amplitudes of the external driving pulses.  In addition to the Taylor expansion in the amplitudes, we also perform a discrete Fourier transform in the phases of the input pulses (Fig.~\ref{fig1}(c)), which will allow us to separate the excitation pathways contributing to the nonlinear signal in Liouville space. This phase expansion plays a role analogous to phase matching via oblique pulses in free space or multimode cavities via phases $\me^{\mi\mathbf{k}_j\cdot\mathbf{r}}$, where the allowed Liouville pathways are selected by wavevector-dependent phase-matching conditions (in addition, for oblique pulses the cavity dispersion must be taken into account) \cite{gelin2009efficient, xiang2018twodimensional, xiang2019manipulating}.

Building on the framework developed in Ref.~\cite{reitz2025nonlinear}, we expand the cavity field and molecular density matrix in terms of both amplitudes and phases of the three pulses as
\begin{subequations}
\label{eq:expansion}
\begin{align}
\alpha  &=\sum_{n,m,l,\vec{v}} \eta_1^n\eta_2^m\eta_3^l\,\me^{-\mi\vec{v}\cdot\vec{\Phi}} \alpha^{(n,m,l)}_{\vec{v}}, \\
\rho  &=\sum_{n,m,l,\vec{v}} \eta_1^n\eta_2^m\eta_3^l\, \me^{-\mi\vec{v}\cdot\vec{\Phi}} \rho^{(n,m,l)}_{\vec{v}},
\end{align}
\end{subequations}
with the vector of phases $\vec{\Phi}=(\Phi_1,\Phi_2,\Phi_3)$, and where
$\vec{v}=(v_1,v_2,v_3)$ selects the corresponding phase harmonics
($v_j\in\mathbb{Z}$), while the superscript $(n,m,l)$ specifies the nonlinear
order in the pulse amplitudes ($n,m,l\in\mathbb{N}_0$).  We remark that the notation $\sum_{\vec{v}}$ in the sums above is shorthand for convenience. More precisely, this is a restricted sum where the possible phase indices (those yielding non-zero contributions) depend on the specific order, i.e., $\vec{v}(n,m,l)$ (see details on phase expansion in SI~\ref{sec:phasedetails}). The phase of each impinging pulse is first imprinted onto the cavity field and subsequently transferred to the molecular density matrix via the light-matter interaction, giving rise to well-defined phase-tagged nonlinear excitation pathways \cite{gelin2005efficient, gelin2009efficient}. Note that both $\alpha$ and $\rho$ are dependent on the phase of the input pulses $\alpha (\vec{\Phi})$, $\rho(\vec{\Phi})$. However, since in practice, in the following we only extract the desired phase components via the inverse transform (see SI~\ref{sec:phasedetails}), the explicit phase dependence is suppressed for notational simplicity. The above expansion can in principle be generalized to an arbitrary number of input pulses.

From the  perturbative expansion, a closed set of equations for the components 
$\alpha^{(n,m,l)}_{\vec{v}}$ and $\rho^{(n,m,l)}_{\vec{v}}$ can be derived (see following section). 
In the following, we restrict the perturbative expansion to first order in each of the three pulses, i.e., we propagate the system up to order $(1,1,1)$, corresponding to a total third order nonlinearity. We will make use of the notation $\alpha^{(n,m,l)}_{\vec{v}\cdot\vec{\Phi}}$ and 
$\rho^{(n,m,l)}_{\vec{v}\cdot\vec{\Phi}}$ in the following, 
so e.g.,\ $\alpha^{(1,1,1)}_{\Phi_1+\Phi_2-\Phi_3}$ for $\vec{v}=(1,1,-1)$, such that the phase 
combination of a specific component can be more easily read from the subscript. At third order, one finds that the only non-zero phase combinations are given by
$\pm \Phi_1 \mp \Phi_2 + \Phi_3$ and $\Phi_1 + \Phi_2 - \Phi_3$, together with the complex-conjugate processes obtained by an overall sign reversal of all phases. The former combinations correspond to the so-called \textit{rephasing} (R, $-\Phi_1 + \Phi_2 + \Phi_3$) and \textit{non-rephasing} (NR, $+\Phi_1 - \Phi_2 + \Phi_3$) contributions, while the latter corresponds to the \textit{double-quantum coherence} (2QC) contribution ($\Phi_1 + \Phi_2 - \Phi_3$), where the first two pulses act twice with the same phase sign and thereby create a double excitation and a coherence between $\lvert g \rangle$ and $\lvert f \rangle$ \cite{mukamel1995principles}. The R and NR contributions constitute single-quantum (1Q) spectra, since during the evolution periods they involve only coherences between adjacent excitation manifolds (with the exception of contributions arising from pulse overlap inside the cavity, as discussed below and in SI \ref{sec:eoms}), and the detected signal therefore reflects single-quantum transition frequencies. The R and NR contributions can be used to disentangle homogeneous from inhomogeneous broadening \cite{jonas2003two}, though in the present work we restrict ourselves to homogeneous broadening only. The phase combinations above are directly analogous to the more familiar phase-matching conditions $\pm\mathbf{k}_1 \mp \mathbf{k}_2 + \mathbf{k}_3$, and $\mathbf{k}_1 + \mathbf{k}_2 - \mathbf{k}_3$ that arise in non-collinear four-wave-mixing geometries \cite{mukamel1995principles}. Both the R and NR signals can be further decomposed into the three standard Liouville-space pathways: stimulated emission (SE), ground-state bleach (GSB), and excited-state absorption (ESA) (see SI~\ref{sec:eoms}, \ref{sec:harmonic} and \ref{sec:contributions} for a more detailed discussion of the individual contributions to the nonlinear response).

\begin{figure}[b]
    \centering
    \includegraphics[width=1.0\columnwidth]{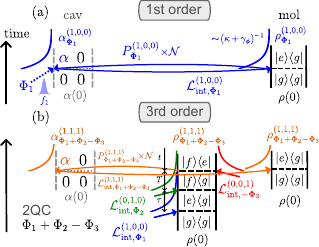}
    \caption{Feynman diagrams for the cavity (left) and molecular density matrix (right) at (a) linear and (b) third order. (a) At linear order, e.g., following the first pulse, the external input field creates a cavity coherence $\alpha_{\Phi_1}^{(1,0,0)}$, which in turn induces a molecular polarization via $\mathcal{L}_{\mathrm{int},\Phi_1}^{(1,0,0)}$. The (collective) polarization $P_{\Phi_1}^{(1,0,0)}$ then acts back on the cavity field, giving rise to the linear polariton response. Both the molecular polarization and the cavity field persist over the extended polaritonic lifetime $\sim(\kappa+\gamma_\phi)^{-1}$ as indicated by the decaying shaded envelopes. (b) At third order, the linear cavity fields generated by the individual pulses combine to create a third-order molecular polarization; shown here is an exemplary 2QC pathway with phase combination $\Phi_1+\Phi_2-\Phi_3$. This collective third-order polarization $P^{(1,1,1)}_{\Phi_1+\Phi_2-\Phi_3}$ generates a third-order cavity field, which then acts back on the zeroth-order molecular density matrix $\rho (0)$ via $\mathcal{L}_{\mathrm{int},\Phi_1+\Phi_2-\Phi_3}^{(1,1,1)}$.}
    \label{fig2}
\end{figure}

\subsection{Liouville space evolution}
\label{sec:Liouville}

The EoM at a certain nonlinear order and phase combination can now be obtained by substituting the expansion from Eqs.~\eqref{eq:expansion} into the mean-field light-matter dynamics [Eqs.~\eqref{eq:meanfieldliouville} and \eqref{eq:cavitymean}] and collecting terms with identical powers in $\eta_j$ as well as matching phase combinations. The projection onto a specific phase
component is performed via an inverse Fourier transform (see
SI~\ref{sec:phasedetails}). Experimentally, this is obtained by cycling
over the pulse phases and isolating the desired phase-dependent contribution
\cite{yuenzhou2014ultrafast}, or alternatively, by isolating a signal at a particular direction through phase matching.  From this, the EoM of the coupled light-matter system at a general nonlinear order $n+m+l>1$ read
\begin{subequations}
\label{eq:perturbative_full}
\begin{align}
\dot{\alpha}^{(n,m,l)}_{\vec{v}}
&=
-\left(\frac{\kappa}{2}+\mi\omega_c\right)
\alpha^{(n,m,l)}_{\vec{v}}
-\mi \mathcal{N}E_0\, P^{(n,m,l)}_{\vec{v}},
\label{eq:perturbative_alpha}
\\[4pt]
\dot{\vec{\rho}}^{(n,m,l)}_{\vec{v}}
&=
-\mi\left( \mathcal{L}_{0}+\mathcal{L}_{\mathcal{D}} \right)
\vec{\rho}^{(n,m,l)}_{\vec{v}}
\notag \\[2pt]
&\quad
-\mi
\sum_{\substack{j,j',j'' = 0 \\[2pt] \vec{u}+\vec{w}=\vec{v}}}^{n,m,l}
\mathcal{L}_{\mathrm{int},\,\vec{u}}^{(j,j',j'')} \,
\vec{\rho}^{(n-j,m-j',l-j'')}_{\vec{w}} ,
\label{eq:perturbative_rho}
\end{align}
\end{subequations}
where the Liouvillians are defined analogously to Eqs.~\eqref{eq:liouvillians}, now expressed in terms of the components of the perturbative expansion. Similarly, the polarization entering Eq.~\eqref{eq:perturbative_alpha} is given by
$P^{(n,m,l)}_{\vec{v}}=\mathrm{Tr}[\hat{\mu}\rho^{(n,m,l)}_{\vec{v}}]$.

The above equations already provide some insight into the structure of the nonlinear response: The nonlinear cavity field in Eq.~\eqref{eq:perturbative_alpha} simply inherits the phase of the molecular polarization of the same order $P^{(n,m,l)}_{\vec{v}}$.  In contrast, the last term in Eq.~\eqref{eq:perturbative_rho} shows that new phase contributions of the density matrix can be generated by the combined action of lower-order cavity field components, encoded in the interaction Liouvillian $\mathcal{L}_{\mathrm{int}}$, and lower-order molecular density matrix components. Importantly, only the terms satisfying the phase-matching condition $\vec{v}=\vec{u}+\vec{w}$ are selected to contribute in the sum. We note that Eq.~\eqref{eq:perturbative_rho} can be formally integrated to yield the full evolution of the molecular state in Liouville space, providing an explicit time-domain expression for each perturbative contribution to the density matrix which can be represented by a double-sided Feynman diagram. Importantly, after Fourier transforming to frequency space, the equations for each fixed perturbative order and phase combination become algebraically linear at that order, with an inhomogeneous source term determined by lower orders. Consequently, each nonlinear contribution can be viewed as the linear polaritonic response to an effective source generated by lower-order dynamics. While the structure of the equation for the density matrix in Eq.~\eqref{eq:perturbative_rho} is similar to free-space spectroscopy, the key difference here is that the interaction Liouvillians are generated by the cavity fields rather than by the external fields which undergo their own dynamics and may themselves depend nonlinearly on the input field~\cite{reitz2025nonlinear}.

The first-order (linear) equations, e.g., for the first pulse, are given by (writing the phase components explicitly here)
\begin{subequations}
\label{eq:linear}
\begin{align}
\dot{\alpha}^{(1,0,0)}_{\Phi_1}
&=-\left(\frac{\kappa}{2}\!+\!\mi\omega_c \right)\alpha^{(1,0,0)}_{\Phi_1}
\!-\!\mi \mathcal{N} E_0 P_{\Phi_1}^{(1,0,0)} \!-\!f_1\me^{-\mi\omega_{\ell,1} t},\\
\dot{\vec{\rho}}^{(1,0,0)}_{\Phi_1}
&=-\mi(\mathcal{L}_0+\mathcal{L}_{\mathcal{D}})\vec{\rho}^{(1,0,0)}_{\Phi_1}
-\mi\mathcal{L}_{\mathrm{int},\Phi_1}^{(1,0,0)}\vec{\rho}^{(0,0,0)}_{\phantom{\Phi_1}},
\end{align}
\end{subequations}
 where analogous expressions hold for the $(0,1,0)$ and $(0,0,1)$ components corresponding to the second and third pulses, respectively. Only at linear orders is the cavity field driven directly by the input pulses and both the cavity field and the molecular density matrix simply inherit the phase
of the driving pulse that generates them. For each of the pulses, these equations reproduce the standard linear polariton response~\cite{koner2024linear}. The Feynman diagrams of the cavity and the molecular density matrix are illustrated in Fig.~\ref{fig2}(a). There are, in addition, the corresponding complex-conjugate diagrams in which the pulse acts on the bra side, generating the opposite-phase cavity field $\alpha^{(0,0,1)*}_{-\Phi_1}$ and coherence $\ket{g}\bra{e}$.  The zeroth-order density vector $\vec{\rho}^{(0,0,0)}=\vec{\rho}(0)$ represents the molecular initial state and carries no phase dependence, as no interactions have occurred yet. In the results presented here, we take for simplicity
$\rho(0)=\ket{g}\!\bra{g}$,
i.e., all molecules initially in the vibrational ground state,
although the formalism readily accommodates arbitrary initial conditions such as thermal states \cite{reitz2025nonlinear}. \\

To make the structure of Eqs.~\eqref{eq:perturbative_full} more concrete, and in particular, to illustrate how the formalism operates at the third-order nonlinearity considered for 2D spectroscopy in the following, let us specialize to the case of the third-order 2QC contribution. For the phase combination $\Phi_1+\Phi_2-\Phi_3$ at order $(1,1,1)$, the coupled light-matter equations are given by
\begin{subequations}
\label{eq:thirdorder_eoms}
\begin{align}
\dot{\alpha}^{(1,1,1)}_{\Phi_1+\Phi_2-\Phi_3}
&=
-\!\left(\frac{\kappa}{2}\!+\!\mi\omega_c\right)
\alpha^{(1,1,1)}_{\Phi_1+\Phi_2-\Phi_3}
\!-\!\mi\mathcal{N} E_0\,P^{(1,1,1)}_{\Phi_1+\Phi_2-\Phi_3},
\label{eq:thirdorder_alpha}
\\[4pt]
\dot{\vec{\rho}}^{(1,1,1)}_{\Phi_1+\Phi_2-\Phi_3}
&=
-\mi\,(\mathcal{L}_0+\mathcal{L}_\mathcal{D})\,
\vec{\rho}^{(1,1,1)}_{\Phi_1+\Phi_2-\Phi_3}
\notag \\[-2pt]
&\quad
-\mi\,\mathcal{L}^{(1,1,1)}_{\mathrm{int},\,\Phi_1+\Phi_2-\Phi_3}\,
\vec{\rho}^{(0,0,0)}
\notag \\[-2pt]
&\quad
-\mi\,\mathcal{L}^{(1,0,0)}_{\mathrm{int},\,\Phi_1}\,
\vec{\rho}^{(0,1,1)}_{\Phi_2-\Phi_3}
\notag \\[-2pt]
&\quad
-\mi\,\mathcal{L}^{(0,1,0)}_{\mathrm{int},\,\Phi_2}\,
\vec{\rho}^{(1,0,1)}_{\Phi_1-\Phi_3}
\notag \\[-2pt]
&\quad
-\mi\,\mathcal{L}^{(0,0,1)}_{\mathrm{int},\,-\Phi_3}\,
\vec{\rho}^{(1,1,0)}_{\Phi_1+\Phi_2} .
\label{eq:thirdorder_rho}
\end{align}
\end{subequations}
In writing Eqs.~\eqref{eq:thirdorder_eoms}, we have used that all second-order cavity field components and therefore interaction Liouvillians vanish: at second order, only coherence between $\ket{g}$ and $\ket{f}$ as well as population in $\ket{e}$ is created but no net polarization, and thus no source term for the cavity field. One of the Feynman diagrams contributing to the last term in Eq.~\eqref{eq:thirdorder_rho} proportional to $\,\mathcal{L}^{(0,0,1)}_{\mathrm{int},\,-\Phi_3}\,
\vec{\rho}^{(1,1,0)}_{\Phi_1+\Phi_2}$  is illustrated in Fig.~\ref{fig2}(b), where the third interaction acts on the coherence prepared by the first two interactions encoded in $\vec{\rho}^{(1,1,0)}_{\Phi_1+\Phi_2}$.  The resulting third-order molecular polarization $P^{(1,1,1)}_{\Phi_1+\Phi_2-\Phi_3}$ generates a third-order cavity field $\alpha^{(1,1,1)}_{\Phi_1+\Phi_2-\Phi_3}$, which then feeds back onto the zeroth-order molecular density matrix, creating coherence between $\ket{g}$ and $\ket{e}$. This is described by the second term in Eq.~\eqref{eq:thirdorder_rho} proportional to $\mathcal{L}^{(1,1,1)}_{\mathrm{int},\,\Phi_1+\Phi_2-\Phi_3}\,
\vec{\rho}^{(0,0,0)}$ and is illustrated by the orange arrows in Fig.~\ref{fig2}(b).

\begin{figure}[b]
    \centering
    \includegraphics[width=1.0\columnwidth]{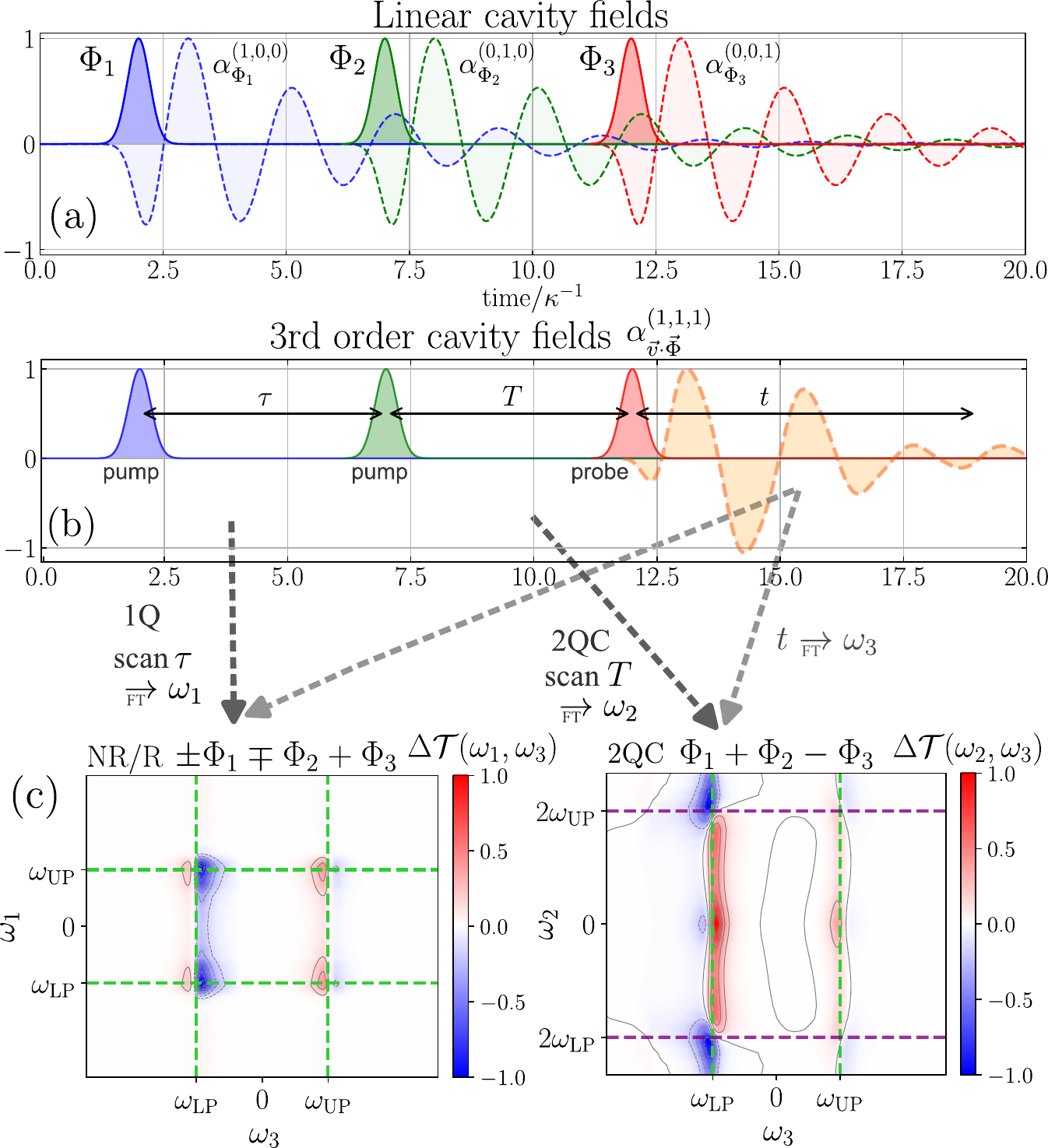}
    \caption{Schematic procedure to obtain phase-cycled 2D cavity spectra: Illustration of (a) linear and (b) nonlinear (3rd order) cavity fields (dashed shaded curves) generated by the Gaussian input pulses (solid shaded curves). Displayed is the real part only, with maxima normalized to unity. The DT signal arises from heterodyne detection, namely the interference between the linear probe (third pulse local oscillator) and $3$rd order cavity fields.  (c) Fourier transform along $\tau$ ($T$) as well as $t$ gives rise to 1Q (2QC) spectra with frequency axes $\omega_1$ ($\omega_2$) vs.~$\omega_3$, respectively. The 1Q spectra can be obtained in the R or NR configurations, corresponding to the phase sequences $\pm\Phi_1\mp\Phi_2+\Phi_3$ (shown is R+NR), while the 2QC pathway corresponds to $\Phi_1+\Phi_2-\Phi_3$. Green dashed lines show single-polariton frequencies, purple dashed lines show double-polariton frequencies (2LP/2UP). Only the real part of the signal is plotted, and each spectrum is normalized by the maximum absolute value of the real part. The spectra are shown in a frame rotating at the central pulse carrier frequency $\omega_{\ell}\equiv\omega_{\ell,j}$. We set $\omega_\ell=\omega_{e}=\omega_c$ (also in all other plots).}
    \label{fig3}
\end{figure}

Importantly, the equations above also contain non-chronological contributions. For instance, the third line in Eq.~\eqref{eq:thirdorder_rho} describes a process in which the interaction associated with the first pulse occurs after the interactions associated with the second and third pulses. Such terms arise because the cavity mode stores and mixes the incoming pulses over the polariton lifetime $\sim (\kappa + \gamma_\phi)^{-1}$ (as indicated by the decaying envelopes in Fig.~\ref{fig2}, also see Fig.~\ref{fig3}(a)), meaning that one cannot impose the assumption of $\delta$-like, non-overlapping pulses as in free-space nonlinear spectroscopy (impulsive limit). From a Fourier-domain perspective, the same effect can be understood as the linear polariton response acting as  a frequency-selective window that filters the external pulses upon entering the cavity \cite{schwennicke2024}. As shown in Figs.~\ref{fig3}(a),(b), the third-order cavity field $(1,1,1)$ is only created once all three pulses have entered
the system, whereas each individual pulse generates a
cavity response already at linear order. The complete integrated solution of the dynamics in Liouville space, including all chronological and non-chronological terms, is given in SI~\ref{sec:fullliouville}. The chronological and non-chronological Feynman diagrams contributing to the 2QC signal of the molecular density matrix are sketched in SI~\ref{sec:eoms}. The explicit EoM for the density matrix elements of the 3LS are also detailed in SI~\ref{sec:eoms}. In total, 24 equations are propagated in the numerical simulations for all phase combinations up to third order.

\subsection{Coherent 2D spectroscopy}
\label{sec:2dspectra}
Having established a systematic framework for the nonlinear light-matter
dynamics and the associated perturbative phase-resolved contributions, we now turn to the computation of the 2D spectra by computing the DT
signals (see schematic procedure in Figs.~\ref{fig3}(a)-(c)), which is the primary experimental observable. The perturbative intracavity fields obtained from the
mean-field dynamics can be related to experimentally observable transmission and reflection
signals via standard cavity input-output theory \cite{gardiner1985input, steck2007quantum}.
Following Ref.~\cite{reitz2025nonlinear}, we define the differential 2D
transmission as the change in transmission when all three pulses are present
relative to the probe-only case,
\begin{align}
\label{eq:dt}
\Delta \mathcal{T}_{\vec{v}\cdot\vec{\Phi}}&(\tau,T,\omega_3)
=
\mathcal{T}^{\mathrm{pump\textrm{-}on}}_{\vec{v}\cdot\vec{\Phi}}(\tau,T,\omega_3)
-
\mathcal{T}^{\mathrm{pump\textrm{-}off}}_{\vec{v}\cdot\vec{\Phi}}(\omega_3)
\notag\\[-2pt]
&\approx
\left(\frac{\kappa}{2}\right)^{\!2}\eta_1\eta_2
\frac{
2\mathrm{Re}\!\left[
\alpha^{(0,0,1)*}_{-\Phi_3}(\omega_3)\,
\alpha^{(3)}_{\vec{v}\cdot\vec{\Phi}}(\tau,T,\omega_3)
\right]
}{f_3(\omega_3)^2},
\end{align}
where we have introduced the short notation
$\alpha^{(3)}\equiv\alpha^{(1,1,1)}$ for the third-order cavity field.
The approximation symbol in Eq.~\eqref{eq:dt} emphasizes that we retain only
the lowest-order nonlinear correction to the transmission, corresponding to 
third order in the input field amplitudes.
The expression above admits a simple and intuitive interpretation: the DT signal arises from interference between the cavity field that is linear in the probe pulse (see Fig.~\ref{fig3}(a)) and the nonlinear cavity field generated by all three pulses (see Fig.~\ref{fig3}(b)). The detection is therefore self-heterodyned, since the third pulse simultaneously acts as probe and local oscillator. Consequently, the measured signal corresponds to a heterodyne-detected four-wave mixing process in which the probe field provides the phase reference $\Phi_3$. Alternatively, an additional fourth pulse can be introduced as a local oscillator, thereby providing an independent phase reference \cite{Hamm_Zanni_2011}. The Fourier transform with respect to the detection
time~$t$ is already performed by the spectrometer detecting the transmission in Eq.~\eqref{eq:dt}, yielding the frequency-resolved
signal in $\omega_3$.

Finally, to obtain the 2D spectra, we Fourier transform the DT signal in Eq.~\eqref{eq:dt} along the
relevant second time variable.  We can then isolate 1Q  non-rephasing (NR) and rephasing (R) contributions by integrating over the excitation time $\tau$ \cite{mukamel2000multidimensional, jonas2003two}
\begin{subequations}
\begin{align}
\Delta\mathcal{T}_{\mathrm{NR}}(\omega_1,T,\omega_3)
&=
\int_{-\infty}^\infty \mathrm{d}\tau\,
\mathrm{e}^{\mi\omega_1\tau}\,
\Delta\mathcal{T}_{\Phi_1-\Phi_2+\Phi_3}(\tau,T,\omega_3),
\label{eq:2Dspectrum_NR}
\\
\Delta\mathcal{T}_{\mathrm{R}}(\omega_1,T,\omega_3)
&=
\int_{-\infty}^\infty \mathrm{d}\tau\,
\mathrm{e}^{\mi\omega_1\tau}\,
\Delta\mathcal{T}_{-\Phi_1+\Phi_2+\Phi_3}(\tau,T,\omega_3),
\label{eq:2Dspectrum_R}
\end{align}
\end{subequations}
as well as the 2QC contribution obtained by integrating over the waiting time $T$
\begin{align}
\Delta\mathcal{T}_{\mathrm{2QC}}(\tau,\omega_2,\omega_3)=
\int_{-\infty}^\infty \mathrm{d}T\,
\mathrm{e}^{\mi\omega_2 T}\,
\Delta\mathcal{T}_{\Phi_1+\Phi_2-\Phi_3}(\tau,T,\omega_3).
\label{eq:2Dspectrum_DQC}
\end{align}
Here, the 1Q signal is obtained by isolating the R and NR contributions with phase signatures $\pm\Phi_1\mp\Phi_2+\Phi_3$, as given in Eqs.~\eqref{eq:2Dspectrum_NR} and \eqref{eq:2Dspectrum_R}. The 2QC signal is selected by the phase combination $\Phi_1+\Phi_2-\Phi_3$, as defined in Eq.~\eqref{eq:2Dspectrum_DQC}. In the following, we present the real part of the complex Fourier-transformed signals, which corresponds to the absorptive component of the 2D spectra, while the imaginary part contains the dispersive contribution. 

Schematic qualitative examples of resulting 2D spectra are shown in Fig.~\ref{fig3}(c). The 1Q spectra display features centered around the single-polariton frequencies $\omega_{\mathrm{UP}}$ and $\omega_{\mathrm{LP}}$ along both frequency axes, reflecting the evolution and detection of single-quantum coherences throughout the sequence. In contrast, the 2QC spectra exhibit characteristic features along the $\omega_2$-axis at energies corresponding to sums of single-polariton energies, $2\omega_{\mathrm{UP}}$, $2\omega_{\mathrm{LP}}$, and $\omega_{\mathrm{UP}}+\omega_{\mathrm{LP}}$, while along the $\omega_3$-axis they display resonances at the single-polariton frequencies $\omega_{\mathrm{UP}}$ and $\omega_{\mathrm{LP}}$. This structure reflects the correlation between double-quantum coherences generated during the second time interval and single-quantum pathways.

\section{Results \& discussion}
\label{sec:results}

Having established the theoretical framework for the nonlinear cavity response and the construction of phase-resolved 2D transmission spectra, we are now in a position to analyze the resulting 1Q and 2QC spectra in detail. We begin by benchmarking
the present formalism through direct comparison with experimentally-measured
1Q spectra of W(CO)$_6$ at short and long waiting times in Sec.~\ref{sec:experimental}. These experiments probe the linear and nonlinear
polariton dynamics in vibrational strong coupling and therefore serve as an ideal point of comparison for the theory. We then discuss the distinct imprints of mechanical ($\Delta$) and electrical anharmonicities ($\delta$) on the 2QC spectra in Sec.~\ref{sec:dqc}. 

\subsection{Comparison with experimental data: 1Q spectra}
\label{sec:experimental}

The experiments were performed on a solution of  W(CO)$_6$, dissolved in hexane where the triply-degenerate carbonyl asymmetric stretch $T_{1u}$ was strongly coupled to the confined infrared modes of a coplanar Fabry-Pérot cavity \cite{xiang2018twodimensional, xiang2019manipulating}. The CO stretching mode exhibits an anharmonic shift $\Delta\approx -15\,\mathrm{cm}^{-1}$, which can be well characterized by nonlinear spectroscopy outside of the cavity \cite{xiang2018twodimensional, pyles2024revisiting}, making this system an ideal benchmark for testing the capabilities of our approach. In addition, the carbonyl stretching band of W(CO)$_6$ in solution is known to be dominated by homogeneous broadening, with only a minor contribution from inhomogeneous (heterogeneous) broadening. This predominantly homogeneous linewidth justifies modeling the ensemble within an effective single-molecule description, in which the vibrational response is represented by a Lorentzian lineshape.  To better reproduce the experimental observations, we do not assume harmonic dephasing when modeling the actual experiment, and instead assign different homogeneous linewidths to the $g\!\leftrightarrow\!e$ and $e\!\leftrightarrow\!f$ transitions. These linewidths, along with all other parameters used in the simulations, are listed in Table~\ref{tab:exp_params}. Since the electrical anharmonicity of W(CO)$_6$ is unknown, we choose a value reported for the carbonyl stretch in a different system \cite{Khalil2003coherence}.

\begin{figure}[t]
    \centering
    \includegraphics[width=1.0\columnwidth]{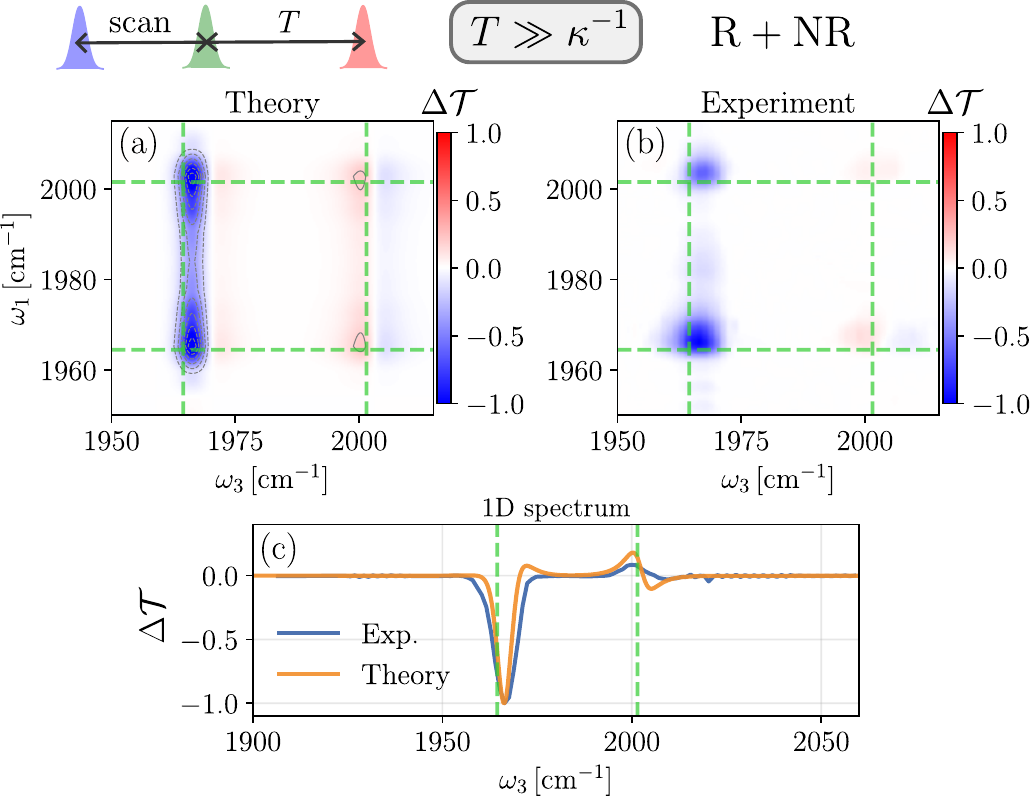}
    \caption{Comparison between (a) theoretical and (b) experimental results for the 2D 1Q spectra at long waiting times $T\gg\kappa^{-1}$, for the sum of NR and R contributions. The waiting time was set to $T=18~\kappa^{-1}$ which corresponds to approximately $55~\mathrm{ps}$. (c) Comparison of the resulting 1D spectra along $\omega_3$, obtained by integrating the 2D spectra over $\omega_1$.  The green dashed lines in all plots indicate the linear polariton frequencies.}
    \label{fig4}
\end{figure}

\begin{figure}[t]
    \centering    \includegraphics[width=1.0\columnwidth]{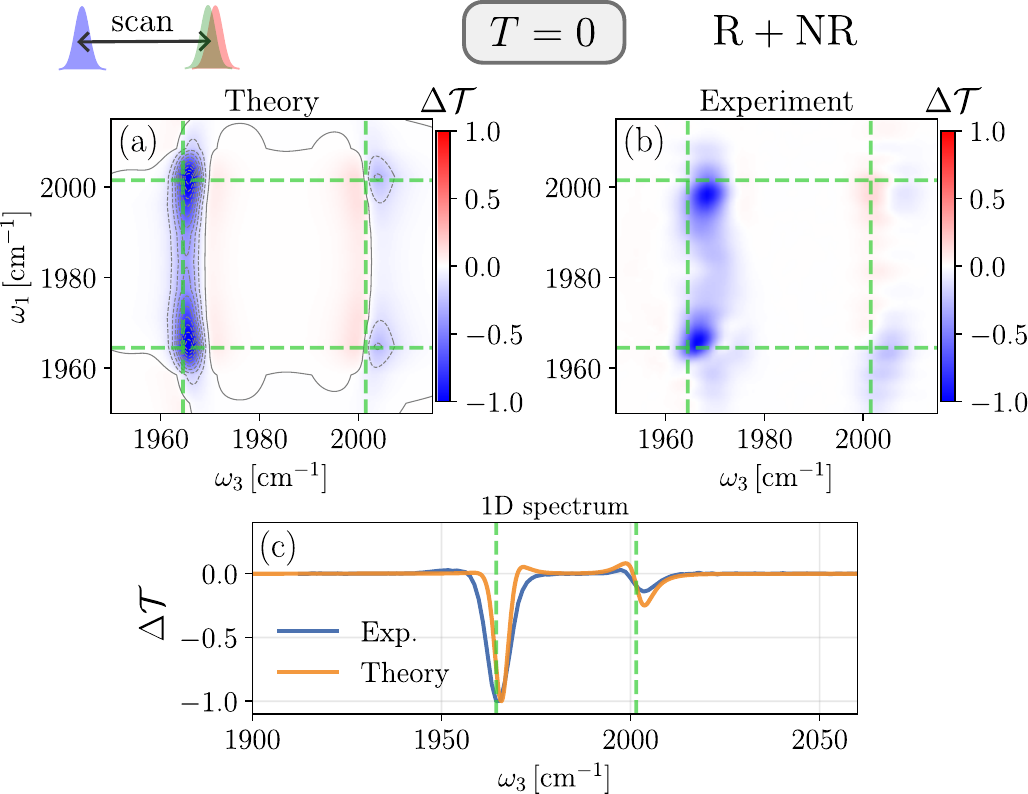}
    \caption{Comparison between (a) theoretical and (b) experimental results for the 2D 1Q spectra and (c) corresponding 1D spectra at short waiting times $T=0$, for the sum of NR and R contributions. To reproduce these results, EID with $\beta=10.0~\mathrm{cm}^{-1}$ was added (see SI \ref{sec:polaritonbleach} for details). All other parameters are the same as in Fig.~\ref{fig4} and Table \ref{tab:exp_params}.}   
    \label{fig4_short}
\end{figure}

\begin{table}[b]
\centering
\caption{Parameters used in the simulations, extracted from Refs.~\cite{xiang2018twodimensional, xiang2019manipulating, ribeiro2018theory}. 
Here, $g_{ge}=\mu_{ge}E_0$ denotes the single-molecule light-matter coupling strength of the fundamental vibrational transition.}
\label{tab:exp_params}
\begin{tabular}{l c}
\hline\hline
Parameter & Value \\
\hline
Collective (linear) light-matter coupling $g_{ge}\sqrt{\mathcal{N}}$ & 18.5 cm$^{-1}$ \\
Fundamental transition frequency $\omega_{e}-\omega_g$ & 1983 cm$^{-1}$ \\
Anharmonic transition frequency $\omega_{f}-\omega_e$ & 1968 cm$^{-1}$ \\
Linewidth of the fundamental $g\leftrightarrow e$ transition $\gamma_{\phi}^{ge}$ & 6 cm$^{-1}$ \\
Linewidth of the $e\leftrightarrow f$ transition $\gamma_{\phi}^{ef}$ & 9 cm$^{-1}$ \\
Cavity linewidth $\kappa$ & 11 cm$^{-1}$ \\
Electrical anharmonicity $\delta$ \cite{Khalil2003coherence} & $-0.25(10)$ \\
\hline\hline
\end{tabular}
\end{table}

A direct comparison between theory and experiment for the 1Q signal at long waiting times, $T\gg\kappa^{-1}$, is shown in Figs.~\ref{fig4}(a) and (b), respectively, where in both cases the plotted quantity is the real part of the sum of NR and R contributions, $\mathrm{Re}[\Delta\mathcal{T}_{\mathrm{NR}}+\Delta\mathcal{T}_{\mathrm{R}}]$. In this long-$T$ regime, the probe pulse predominantly detects dephased, stationary populations created by the two pump pulses. Here and in the following, all frequencies are reported in wavenumber units (cm$^{-1}$) to facilitate direct comparison with the experimental spectra. The corresponding projection onto the 1D spectrum along the detection frequency $\omega_3$ is displayed in Fig.~\ref{fig4}(c), demonstrating good overall agreement between theory and experiment.

The 1Q spectra exhibit several characteristic features. The pronounced blue (negative) region centered around the lower polariton frequency on the $\omega_3$-axis arises predominantly from ESA, reflecting population in the singly excited manifold that can be promoted to higher-lying states. This ESA contribution dominates the nonlinear response in amplitude. Note that a small dark state peak is observed in the experimental 2D spectrum around $\omega_1\approx\omega_e$, $\omega_3\approx\omega_{\mathrm{LP}}$, as well as in the linear reflection and absorption spectra \cite{xiang2018twodimensional}. We do not reproduce this feature here because our model neglects inhomogeneous broadening, which mixes bright and dark states and therefore transfers photonic weight to the dark states. On the higher-frequency side, the weaker alternating red–blue (positive–negative) structure originates from pump-induced population that renormalizes the light-matter coupling. This contribution is associated with GSB and SE (see SI~\ref{sec:contributions}) and effectively reduces the Rabi splitting, producing a derivative-like lineshape along $\omega_3$ in $\Delta \mathcal{T}$.

An important question is whether the theory can also reproduce the short-time dynamics, $T < \kappa^{-1}$, where coherent effects are expected to play a decisive role. For the cavity linewidth considered here, $\kappa = 11~\mathrm{cm}^{-1}$, this corresponds to delay times shorter than approximately $3~\mathrm{ps}$. In the experiment of Ref.~\cite{xiang2019manipulating}, a so-called polariton bleach effect was observed at such short waiting times $T$, manifesting as reduced transmission (enhanced absorption) around both the lower- and upper-polariton frequencies. That work also showed that this effect does not originate from a simple population-induced reduction in oscillator strength and instead can be reproduced phenomenologically by introducing a pump-induced increase of the linewidth of the fundamental vibrational transition. The effect has however remained elusive from a microscopic viewpoint, seemingly necessitating a different interpretation than the usual Rabi splitting contraction~\cite{xiang2019manipulating}. Within the present framework, this behavior can be accounted for by introducing excitation-induced dephasing (EID), such that the dephasing rate depends on the excited-state population, $\gamma_\phi(\rho_{ee})$. Such terms are, e.g., well known to play an important role in semiconductor physics, where a high density of excitons leads to enhanced dephasing \cite{fehrenbach1982transient, wang1993transient,jahnke1996excitonic, takemura2015dephasing}. To leading order, the strength of the EID is controlled by a parameter $\beta$, defined as the derivative of the dephasing rate with respect to the excited-state population, i.e., \ $\beta = \left.\partial \gamma_\phi / \partial \rho_{ee}\right|_{\rho_{ee}=0}$ (see SI~\ref{sec:polaritonbleach} for details). In the perturbative treatment employed here, the second-order populations renormalize the third-order coherences of the fundamental vibrational mode through the population dependence of the decay rate, thereby producing the observed short-time bleach response. Fig.~\ref{fig4_short} presents a comparison between experimental and theoretical 2D and 1D spectra, including EID with $\beta = 10~\mathrm{cm}^{-1}$ at $T = 0$, and shows that this mechanism reproduces the absorptive feature near the upper polariton frequency. Importantly, including these terms also preserves the contraction observed around the upper polariton frequency in the long-time limit (see SI~\ref{sec:polaritonbleach}), thereby providing a consistent description of the experiment in both regimes. Moreover, this term also correctly captures the dependence on concentration and cavity length observed in Ref.~\cite{xiang2019manipulating}, namely the decrease of the polariton bleach signal with increasing concentration and cavity length (see SI \ref{sec:polaritonbleach}). In solid state systems, EID emerges from Coulomb interactions between photoinduced electron and hole densities \cite{jahnke1996excitonic, wang1993transient}. EID in W(CO)$_6$ could emerge from the fact that the single vibrational mode in our model in reality corresponds to a triply-degenerate set that can feature weak anharmonic couplings among the different modes; this would imply that EID would be a weaker effect for vibrational modes that are not degenerate, although a more explicit model must be constructed to properly validate this. 

\begin{figure*}[t]
    \centering
    \includegraphics[width=1.0\textwidth]{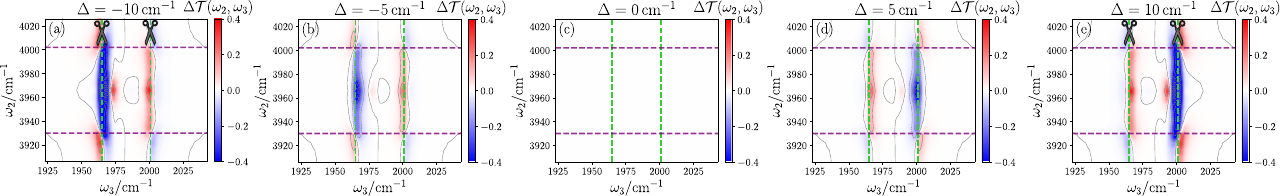}
    \caption{2QC spectra as extracted from the phase combination $\Phi_1+\Phi_2-\Phi_3$ with varying degree of mechanical anharmonicity
(a) $\Delta = -10\,\mathrm{cm}^{-1}$,
(b) $\Delta = -5\,\mathrm{cm}^{-1}$,
(c) $\Delta = 0\,\mathrm{cm}^{-1}$ (harm.~limit),
(d) $\Delta = 5\,\mathrm{cm}^{-1}$,
(e) $\Delta = 10\,\mathrm{cm}^{-1}$. Green vertical lines indicate the single-polariton frequencies, while purple horizontal lines indicate the double-polariton frequencies. To isolate the effect of pure mechanical anharmonicity, we set the electrical anharmonicity to zero ($\delta=0$) and assume pure harmonic dephasing with $\gamma_{\phi}=6~\mathrm{cm}^{-1}$ ($\gamma_\phi=\gamma_{\phi}^{ge}=\gamma_\phi^{ef}$). All other parameters are identical to those listed in Table~\ref{tab:exp_params}. We chose an excitation time of $\tau=26\,\mathrm{fs}$. A movie of the spectra as a function of different excitation times for $\Delta=-10\,\mathrm{cm}^{-1}$ is provided as Supplementary Material. The scissor symbols in (a) and (e) indicate the cuts taken along $\omega_2$ shown in Fig.~\ref{fig7}.}
    \label{fig5}
\end{figure*}

\begin{figure}[t]
    \centering
    \includegraphics[width=0.95\linewidth]{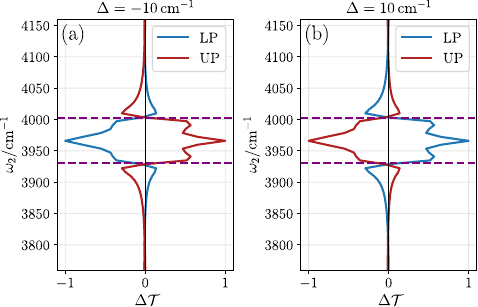}
    \caption{
    Vertical cuts of the 2QC spectra in Fig.~\ref{fig5} at fixed emission frequencies $\omega_3$ corresponding to the lower polariton $\omega_{\mathrm{LP}}$ and upper polariton $\omega_{\mathrm{UP}}$, shown for (a) $\Delta=-10\,\mathrm{cm}^{-1}$ and (b) $\Delta=10\,\mathrm{cm}^{-1}$. Each curve is normalized to its maximum absolute value. The dashed purple horizontal lines indicate the harmonic double-polariton energies.}
    \label{fig7}
\end{figure}

While the agreement between theory and experiment is generally good, some discrepancies remain. For instance, in Fig.~\ref{fig4}(c), the theory seems to overestimate the derivative feature around the upper polariton frequency and generally underestimates the linewidths in the differential spectra. This may have several reasons: First, the spatial structure of the cavity mode can lead to an effective inhomogeneous distribution of coupling strengths which is not taken into account here. Moreover, the presented theoretical results capture only the leading (third-order) correction to the linear response, whereas the experiment may involve higher-order population effects and a larger steady-state excited-state fraction. Also,  the experiment is performed using phase matching with obliquely incident pulses; the angular dispersion of the cavity could therefore introduce additional broadening or frequency shifts not captured in the current model. Finally, additional loss or dephasing channels, including coupling to other modes, are not explicitly included here and could further contribute to the observed disagreement \cite{xiang2019state}.

\subsection{2QC spectroscopy}
\label{sec:dqc}

We now move to 2QC spectroscopy, obtained under the phase-matching condition $\Phi_1+\Phi_2-\Phi_3$ and by Fourier transforming over the waiting-time interval $T$. Fig.~\ref{fig5} shows the 2QC DT spectra 
$\Delta \mathcal{T}_{\mathrm{2QC}}(\omega_2,\omega_3)$ of the cavity-molecule polariton system for various matter anharmonicities $\Delta$. In constructing these spectra, we choose parameter values 
inspired by, and of comparable magnitude to, those used in the previous section for 
W(CO)$_6$, in order to remain in a physically realistic vibrational strong-coupling regime. While positive anharmonicities ($\Delta > 0$) are not typically realized for vibrational modes, which generally exhibit negative (red-shifting) anharmonicity (with exceptions reported, e.g.,  for some hydrogen-bonded water systems \cite{dahms2017large, Dereka2021crossover}), we nevertheless include this case to illustrate how the sign of the anharmonicity modifies the double-quantum response. The $\omega_3$-axis resolves the probe frequency and therefore highlights the 
single-polariton resonances (green dashed lines), while the $\omega_2$-axis tracks the 
double-quantum evolution and becomes resonant at the two-polariton 
(double-polariton) energies (purple dashed lines). In the harmonic limit ($\Delta = 0$), 
the signal vanishes, consistent with the absence of nonlinearity and the 
cancellation of all nonlinear pathways in an equally-spaced ladder (see SI~\ref{sec:harmonic}). 

\begin{figure*}[t]
    \centering
    \includegraphics[width=1.0\textwidth]{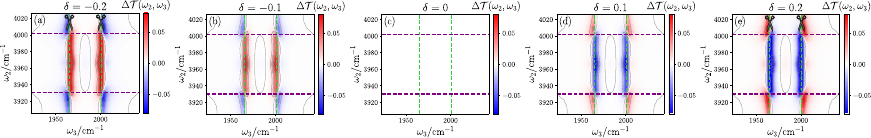}
    \caption{2QC spectra as extracted from the phase combination $\Phi_1+\Phi_2-\Phi_3$ with varying degree of electrical anharmonicity
(a) $\delta = -0.2$,
(b) $\delta = -0.1$,
(c) $\delta = 0$ (harm.~limit),
(d) $\delta = 0.1$,
(e) $\delta = 0.2$. 
Green vertical lines indicate the single-polariton frequencies, while purple horizontal lines indicate the double-polariton frequencies. To isolate the effect of pure electrical anharmonicity, we set the mechanical anharmonicity to zero ($\Delta=0$) and assume pure harmonic dephasing with $\gamma_{\phi}=6~\mathrm{cm}^{-1}$ ($\gamma_\phi=\gamma_{\phi}^{ge}=\gamma_\phi^{ef}$). All other parameters are identical to those listed in Table~\ref{tab:exp_params} and Fig.~\ref{fig5}. The scissor symbols in (a) and (e) indicate the cuts taken along $\omega_2$ shown in Fig.~\ref{fig8}.}
    \label{fig6}
\end{figure*}

\begin{figure}[t]
    \centering
    \includegraphics[width=0.95\linewidth]{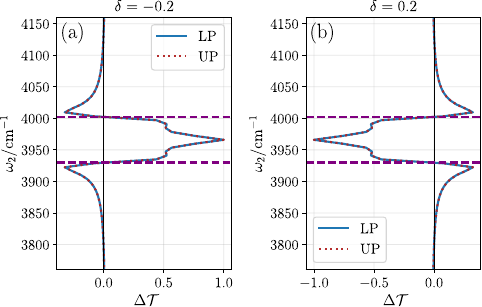}
    \caption{
    Vertical cuts of the 2QC spectra in Fig.~\ref{fig6} at fixed emission frequencies $\omega_3$ corresponding to the lower polariton $\omega_{\mathrm{LP}}$ and upper polariton $\omega_{\mathrm{UP}}$, shown for (a) $\delta=-0.2$ and (b) $\delta=0.2$. Each curve is normalized to its maximum absolute value. The dashed purple horizontal lines indicate the harmonic double-polariton energies.}
    \label{fig8}
\end{figure}

Introducing finite anharmonicity lifts this cancellation by shifting the two-excitation manifold relative to the simple sum of single-polariton energies, thereby generating a finite third-order response. In the perturbative third-order, this shifting of energies appears as derivative-like lineshape features. Note that, although the anharmonicity seen by a single photon vanishes in the $\mathcal{N}\to\infty$ limit \cite{campos2021generalization}, we still observe a considerable anharmonicity because many photons are present inside the cavity. The emission frequencies along $\omega_3$ remain fixed at the LP and UP energies, while the spectral restructuring occurs primarily along $\omega_2$, demonstrating that $\Delta$ acts on the two-excitation eigenstates rather than on the single-polariton branches. The redistribution of intensity is localized near the double-excitation energies, confirming that the observed nonlinear response originates from modifications of the second-manifold polaritons.

This behavior is isolated in Fig.~\ref{fig7} by selecting emission at the LP and UP single-polariton frequencies, thereby resolving how individual two-excitation eigenstates project onto the lower and upper polariton components. In physical terms, these cuts separate the contributions of double-polariton states according to their polaritonic character, allowing one to directly track how anharmonicity redistributes lower- and upper-polaritonic content within the second excitation manifold. For $\Delta<0$, the LP channel exhibits an expansion of the dominant double-excitation resonances, while the UP channel shows a contraction, this trend reverses for $\Delta>0$. Concomitantly, spectral weight shifts between the LP and UP emission pathways. This behavior is consistent with level repulsion within the coupled second-excitation manifold: tuning $\Delta$ redistributes the avoided-crossing structure, such that one adjacent spacing increases while the other decreases. The opposite LP/UP response reflects how this eigenvalue reshuffling projects differently onto the single polariton emission channels. A full movie of the 2QC signal for a mechanical anharmonicity of $\Delta=-10\,\mathrm{cm}^{-1}$ as a function of excitation time $\tau$ is provided as Supplementary Material.

The 2QC spectra for varying electrical anharmonicity $\delta$ at vanishing mechanical anharmonicity ($\Delta=0$), thereby isolating dipole-induced nonlinearities, are displayed in Fig.~\ref{fig6}. In contrast to mechanical anharmonicity, which reshuffles the eigenenergies of the two-excitation manifold and redistributes polaritonic character asymmetrically among its eigenstates, electrical anharmonicity modifies the effective light-matter coupling in the double-excitation sector without shifting the underlying harmonic energy ladder of the bare molecule. As a consequence, it changes the magnitude of the two-polariton splitting while preserving symmetry between the LP and UP emission channels along $\omega_3$.

For $\delta<0$, the effective light-matter coupling in the double-excitation sector is reduced, bringing the double-polariton frequencies closer to the spectral center and thereby contracting the two-polariton manifold along the $\omega_2$-axis (see cuts along single-polariton frequencies in Fig.~\ref{fig8}(a)). Within the perturbative third-order spectra, this contraction is reflected in derivative-like line shapes whose sign encodes the underlying displacement. In addition, for $\delta<0$ one observes enhanced transmission around $\omega_{\mathrm{LP}}+\omega_{\mathrm{UP}}$, which is likewise evident in the cuts. For $\delta>0$, by contrast, the effective coupling is enhanced, leading to an outward expansion of the double-excitation manifold and, correspondingly, to derivative features of opposite sign (see cuts in Fig.~\ref{fig8}(b)). In this case, the transmission around $\omega_{\mathrm{LP}}+\omega_{\mathrm{UP}}$ is reduced, corresponding to enhanced absorption. Moreover, as becomes apparent in Fig.~\ref{fig6}, increasing the anharmonicity does not produce a larger apparent shift of the resonances in the perturbative approach, but instead amplifies the magnitude of the nonlinear response. In the harmonic limit, $\delta=0$, the nonlinear pathways cancel exactly again and the 2QC signal vanishes.

\section{Conclusions and outlook}
\label{sec:conclusion}

We have presented a general and computationally efficient approach to compute multidimensional polariton spectra, based on a semiclassical evolution of the coupled light-matter system. The formalism enables the straightforward construction of phase-cycled multidimensional spectra from the underlying nonlinear signal components.  We have employed a simplified molecular model consisting of an anharmonic 3LS subject to pure dephasing.

We have compared the method against experimental 1Q spectra at short and long waiting times and found overall good agreement, showing that the polariton bleach can be explained by the introduction of excitation-induced dephasing, leading to broadening of the fundamental vibrational transition. Future work will investigate the origin of this effect and, in particular, whether it arises from anharmonic couplings within the triply degenerate set of vibrational modes. We have also discussed the imprint of both mechanical and electrical anharmonicities on 2QC spectra. While our description captures the essential nonlinear polaritonic response, more sophisticated models\textemdash for instance including structured or non-Markovian baths, vibrational environments, inhomogeneous broadening \cite{liu2025unlocking, yin2025overcoming}, or multimode cavities\textemdash could also  be incorporated within the same framework. Such extensions are expected to provide closer quantitative agreement with the experimentally-observed multidimensional spectra. More broadly, we expect the present method to be applicable to most current experiments probing ultrafast nonlinear polariton dynamics. A straightforward and physically relevant extension would be the inclusion of multiple molecular subensembles, such as donor-acceptor systems, coupled to the same cavity mode. This would enable the modeling of ultrafast spectroscopy of polariton-mediated energy transfer \cite{xiang2020intermolecular, mewes2020energy}.

An important open question concerns the connection of our semiclassical formalism to descriptions in terms of polariton excitation manifolds of the Tavis-Cummings model \cite{delpo2020polariton, fassioli2021femtosecond, autry2020excitation}. Understanding how, and under what conditions, the semiclassical dynamics reproduces or deviates from the quantum manifold picture will be addressed in a future study. Further, correlations between light and matter, which are absent at the current semiclassical level, could be systematically reincorporated through controlled expansions in powers of $1/\mathcal{N}$, providing a pathway to include entanglement and quantum optical effects beyond mean-field theory \cite{perezsanchez2025cute, fowlerwright2023determining}. In parallel, we are currently pursuing the integration of this approach with finite-difference time-domain (FDTD) methods, thereby enabling the description of ultrafast nonlinear polariton spectroscopy in realistic, arbitrarily structured electromagnetic environments \cite{zhou2024simulating}.\\


\section{Acknowledgments}

 This research was primarily supported by the Air Force Office
of Scientific Research (AFOSR) through the Multi-University Research Initiative (MURI)
program no.~FA9550-22-1-0317. We thank Arghadip Koner for help in preparing the schematic of the cavity shown in Fig.~\ref{schem}. \\

\bibliography{multi}

\clearpage
\onecolumngrid

\setcounter{section}{0}
\setcounter{equation}{0}
\setcounter{figure}{0}
\setcounter{table}{0}

\renewcommand{\thesection}{S\arabic{section}}
\renewcommand{\theequation}{S\arabic{equation}}
\renewcommand{\thefigure}{S\arabic{figure}}
\renewcommand{\thetable}{S\arabic{table}}

\makeatletter
\newcommand{\supplementarytocfile}{stoc}
\let\oldsection\section
\renewcommand{\section}{%
  \@ifstar{\oldsection*}{\suppsection}%
}
\newcommand{\suppsection}[1]{%
  \oldsection{#1}%
  \addcontentsline{\supplementarytocfile}{section}{\protect\numberline{\thesection}#1}%
}
\makeatother

\begin{center}
\large
Supplementary Information for\\[4pt]
\textbf{``Multidimensional semiclassical single- and double-quantum spectroscopy of anharmonic molecular polaritons''}
\end{center}

\phantomsection
\vspace{0.5em}
\begin{center}
\large\bfseries Contents
\end{center}
\vspace{0.25em}

{
\setcounter{tocdepth}{2}

\parskip=0pt
\itemsep=0pt
\parsep=0pt

\makeatletter
\def\l@section#1#2{\addpenalty\@secpenalty\addvspace{0.2em}%
  \setlength\@tempdima{1.5em}%
  \begingroup
    \parindent \z@ \rightskip \@pnumwidth
    \parfillskip -\@pnumwidth
    \leavevmode 
    \advance\leftskip\@tempdima
    \hskip -\leftskip
    #1\nobreak\hfil \nobreak\hb@xt@\@pnumwidth{\hss #2}\par
  \endgroup
}

\def\l@subsection#1#2{%
  \setlength\@tempdima{2.8em}%
  \begingroup
    \parindent \z@ \rightskip \@pnumwidth
    \parfillskip -\@pnumwidth
    \leavevmode
    \advance\leftskip\@tempdima
    \hskip -\leftskip
    #1\nobreak\hfil \nobreak\hb@xt@\@pnumwidth{\hss #2}\par
  \endgroup
}

\addtocontents{\supplementarytocfile}{\protect\setcounter{tocdepth}{2}}
\@starttoc{\supplementarytocfile}
\makeatother
}

\section{Liouville space evolution}
\label{sec:fullliouville}

The integrated solution for the molecular density matrix in terms of the free molecular propagator (Green's function) $\theta (t-t')\me^{-\mi\mathcal{L}_m(t-t')}=\mathcal{G}(t-t')$ where $\mathcal{L}_m=\mathcal{L}_0+\mathcal{L}_\mathcal{D}$ reads
\begin{align}
\nonumber{\vec\rho}^{(1,1,1)} (t)& =
-\mi\int_0^t \td t_1\, \mathcal{G}(t-t_1)\mathcal{L}\ts{int}^{(1,1,1)}\vec\rho (0),\\\nonumber
&\quad
+(-\mi)^2\int_0^t \td t_2 \int_0^{t_2}\td t_1\,\mathcal{G}(t-t_2)\mathcal{L}\ts{int}^{(1,1,0)}(t_2)\mathcal{G}(t_2-t_1)\mathcal{L}\ts{int}^{(0,0,1)}(t_1)\vec\rho (0)\\\nonumber
&\quad
+(-\mi)^2\int_0^t \td t_2 \int_0^{t_2}\td t_1\,\mathcal{G}(t-t_2)\mathcal{L}\ts{int}^{(1,0,1)}(t_2)\mathcal{G}(t_2-t_1)\mathcal{L}\ts{int}^{(0,1,0)}(t_1)\vec\rho (0)\\\nonumber
&\quad
+(-\mi)^2\int_0^t \td t_2 \int_0^{t_2}\td t_1\,\mathcal{G}(t-t_2)\mathcal{L}\ts{int}^{(0,1,1)}(t_2)\mathcal{G}(t_2-t_1)\mathcal{L}\ts{int}^{(1,0,0)}(t_1)\vec\rho (0)\\\nonumber
&\quad
+(-\mi)^2\int_0^t \td t_2 \int_0^{t_2}\td t_1\,\mathcal{G}(t-t_2)\mathcal{L}\ts{int}^{(1,0,0)}(t_2)\mathcal{G}(t_2-t_1)\mathcal{L}\ts{int}^{(0,1,1)}(t_1)\vec\rho (0)\\\nonumber
&\quad
+(-\mi)^2\int_0^t \td t_2 \int_0^{t_2}\td t_1\,\mathcal{G}(t-t_2)\mathcal{L}\ts{int}^{(0,1,0)}(t_2)\mathcal{G}(t_2-t_1)\mathcal{L}\ts{int}^{(1,0,1)}(t_1)\vec\rho (0)\\\nonumber
&\quad
+(-\mi)^2\int_0^t \td t_2 \int_0^{t_2}\td t_1\,\mathcal{G}(t-t_2)\mathcal{L}\ts{int}^{(0,0,1)}(t_2)\mathcal{G}(t_2-t_1)\mathcal{L}\ts{int}^{(1,1,0)}(t_1)\vec\rho (0)\\\nonumber
&\quad
+(-\mi)^3\int_0^t \td t_3 \int_0^{t_3}\td t_2 \int_0^{t_2}\td t_1\,\mathcal{G}(t-t_3)\mathcal{L}\ts{int}^{(1,0,0)}(t_3)\mathcal{G}(t_3-t_2)\mathcal{L}\ts{int}^{(0,1,0)}(t_2)\mathcal{G}(t_2-t_1)\mathcal{L}\ts{int}^{(0,0,1)}(t_1)\vec\rho (0)\\\nonumber
&\quad
+(-\mi)^3\int_0^t \td t_3 \int_0^{t_3}\td t_2 \int_0^{t_2}\td t_1\,\mathcal{G}(t-t_3)\mathcal{L}\ts{int}^{(1,0,0)}(t_3)\mathcal{G}(t_3-t_2)\mathcal{L}\ts{int}^{(0,0,1)}(t_2)\mathcal{G}(t_2-t_1)\mathcal{L}\ts{int}^{(0,1,0)}(t_1)\vec\rho (0)\\\nonumber
&\quad
+(-\mi)^3\int_0^t \td t_3 \int_0^{t_3}\td t_2 \int_0^{t_2}\td t_1\,\mathcal{G}(t-t_3)\mathcal{L}\ts{int}^{(0,1,0)}(t_3)\mathcal{G}(t_3-t_2)\mathcal{L}\ts{int}^{(1,0,0)}(t_2)\mathcal{G}(t_2-t_1)\mathcal{L}\ts{int}^{(0,0,1)}(t_1)\vec\rho (0)\\\nonumber
&\quad
+(-\mi)^3\int_0^t \td t_3 \int_0^{t_3}\td t_2 \int_0^{t_2}\td t_1\,\mathcal{G}(t-t_3)\mathcal{L}\ts{int}^{(0,1,0)}(t_3)\mathcal{G}(t_3-t_2)\mathcal{L}\ts{int}^{(0,0,1)}(t_2)\mathcal{G}(t_2-t_1)\mathcal{L}\ts{int}^{(1,0,0)}(t_1)\vec\rho (0)\\\nonumber
&\quad
+(-\mi)^3\int_0^t \td t_3 \int_0^{t_3}\td t_2 \int_0^{t_2}\td t_1\,\mathcal{G}(t-t_3)\mathcal{L}\ts{int}^{(0,0,1)}(t_3)\mathcal{G}(t_3-t_2)\mathcal{L}\ts{int}^{(1,0,0)}(t_2)\mathcal{G}(t_2-t_1)\mathcal{L}\ts{int}^{(0,1,0)}(t_1)\vec\rho (0)\\
&\quad
+(-\mi)^3\int_0^t \td t_3 \int_0^{t_3}\td t_2 \int_0^{t_2}\td t_1\,\mathcal{G}(t-t_3)\mathcal{L}\ts{int}^{(0,0,1)}(t_3)\mathcal{G}(t_3-t_2)\mathcal{L}\ts{int}^{(0,1,0)}(t_2)\mathcal{G}(t_2-t_1)\mathcal{L}\ts{int}^{(1,0,0)}(t_1)\vec\rho (0).
\end{align}
For simplicity, we have dropped the  subscripts describing the phase indices, as the structure of the integrated solution is identical for all phase combinations. The last line would correspond to the free-space result  with spatially-separated ($\delta$-like) pulses, corresponding to sequential application of the three pulses.

Making use of the fact that for the 3LS considered here all second-order fields are vanishing, leaves only the integrals
\begin{align}
\label{eq:liouville}
\nonumber{\vec\rho}^{(1,1,1)} (t)& =
-\mi\int_0^t \td t_1\, \mathcal{G}(t-t_1)\mathcal{L}\ts{int}^{(1,1,1)}\vec\rho (0),\\\nonumber
&\quad
(-\mi)^3\int_0^t \td t_3 \int_0^{t_3}\td t_2 \int_0^{t_2}\td t_1\,\mathcal{G}(t-t_3)\mathcal{L}\ts{int}^{(1,0,0)}(t_3)\mathcal{G}(t_3-t_2)\mathcal{L}\ts{int}^{(0,1,0)}(t_2)\mathcal{G}(t_2-t_1)\mathcal{L}\ts{int}^{(0,0,1)}(t_1)\vec\rho (0)\\\nonumber
&\quad
(-\mi)^3\int_0^t \td t_3 \int_0^{t_3}\td t_2 \int_0^{t_2}\td t_1\,\mathcal{G}(t-t_3)\mathcal{L}\ts{int}^{(1,0,0)}(t_3)\mathcal{G}(t_3-t_2)\mathcal{L}\ts{int}^{(0,0,1)}(t_2)\mathcal{G}(t_2-t_1)\mathcal{L}\ts{int}^{(0,1,0)}(t_1)\vec\rho (0)\\\nonumber
&\quad
(-\mi)^3\int_0^t \td t_3 \int_0^{t_3}\td t_2 \int_0^{t_2}\td t_1\,\mathcal{G}(t-t_3)\mathcal{L}\ts{int}^{(0,1,0)}(t_3)\mathcal{G}(t_3-t_2)\mathcal{L}\ts{int}^{(1,0,0)}(t_2)\mathcal{G}(t_2-t_1)\mathcal{L}\ts{int}^{(0,0,1)}(t_1)\vec\rho (0)\\\nonumber
&\quad
(-\mi)^3\int_0^t \td t_3 \int_0^{t_3}\td t_2 \int_0^{t_2}\td t_1\,\mathcal{G}(t-t_3)\mathcal{L}\ts{int}^{(0,1,0)}(t_3)\mathcal{G}(t_3-t_2)\mathcal{L}\ts{int}^{(0,0,1)}(t_2)\mathcal{G}(t_2-t_1)\mathcal{L}\ts{int}^{(1,0,0)}(t_1)\vec\rho (0)\\\nonumber
&\quad
(-\mi)^3\int_0^t \td t_3 \int_0^{t_3}\td t_2 \int_0^{t_2}\td t_1\,\mathcal{G}(t-t_3)\mathcal{L}\ts{int}^{(0,0,1)}(t_3)\mathcal{G}(t_3-t_2)\mathcal{L}\ts{int}^{(1,0,0)}(t_2)\mathcal{G}(t_2-t_1)\mathcal{L}\ts{int}^{(0,1,0)}(t_1)\vec\rho (0)\\
&\quad
(-\mi)^3\int_0^t \td t_3 \int_0^{t_3}\td t_2 \int_0^{t_2}\td t_1\,\mathcal{G}(t-t_3)\mathcal{L}\ts{int}^{(0,0,1)}(t_3)\mathcal{G}(t_3-t_2)\mathcal{L}\ts{int}^{(0,1,0)}(t_2)\mathcal{G}(t_2-t_1)\mathcal{L}\ts{int}^{(1,0,0)}(t_1)\vec\rho (0).
\end{align}
In short notation, we can write this as
\begin{align}
\nonumber{\vec\rho}^{(1,1,1)} (t)& =
-\mi\int_0^t \td t_1\, \mathcal{G}(t-t_1)\mathcal{L}\ts{int}^{(1,1,1)}\vec\rho (0),\\
&\quad
(-\mi)^3\sum_{\sigma\in S_3}\int_0^t \td t_3 \int_0^{t_3}\td t_2 \int_0^{t_2}\td t_1\,\mathcal{G}(t-t_3)\mathcal{L}_{\mathrm{int}}^{\s_3}(t_3)\mathcal{G}(t_3-t_2)\mathcal{L}_{\mathrm{int}}^{\s_2}(t_2)\mathcal{G}(t_2-t_1)\mathcal{L}^{\s_1}_{\mathrm{int}}(t_1)\vec{\rho}(0),
\end{align}
where $S_3$ is the set of permutations of $\{(1,0,0), (0,1,0), (0,0,1)\}$ and $\{\s_1, \s_2, \s_3\}$ denotes one such ordered triple. In practice, for the computation of the nonlinear response, we however do not evaluate these nested time integrals explicitly. Instead, we solve the corresponding EoM for the density matrix order by order (see Sec.~\ref{sec:eoms}).

\section{Details on phase expansion}
\label{sec:phasedetails}

In this section, we provide some additional details on the expansion in the pulse phases. For a contribution of order $(n,m,\ell)$ in the three field amplitudes, we define the multi-index 
$\vec n=(n_1,n_2,n_3)\equiv(n,m,\ell)$ and the corresponding phase-index vector 
$\vec v (\vec{n})=(v_1(n_1),v_2(n_2),v_3(n_3))$, where each component $v_i$ is associated with the $i$th field and depends only on $n_i$.

The allowed phase indices (yielding non-zero contributions) are given by
\begin{align}
v_i(n_i)=\sum_{j=1}^{n_i}s_j^{(i)}, \qquad s_j^{(i)}\in \{+1,-1\},
\end{align}
so that
\begin{align}
v_i(n_i) \in \{-n_i,-n_i+2,\hdots, n_i-2, n_i\},
\end{align}
i.e., $v_i$ has the same parity as $n_i$ and changes in steps of two.

Thus, for fixed $(n_1,n_2,n_3)$, the summation over phase indices reads
\begin{align}
\sum_{\vec{v}}\hdots = \sumbyparity{v_1}{-n_1}{n_1}\sumbyparity{v_2}{-n_2}{n_2}\sumbyparity{v_3}{-n_3}{n_3}\hdots,
\end{align}
where $\sumbyparitysymb{}$ (``parity") indicates summation in steps of two.

The inverse transform that extracts a given phase component $\alpha_{\vec{v}}^{(n,m,\ell)}$ from the phase-dependent signal $\alpha^{(n,m,\ell)}(\vec\Phi)$ is
\begin{align}
\alpha_{\vec{v}}^{(n,m,\ell)}=\frac{1}{(2\pi)^3}\int_0^{2\pi}\td^3\Phi\, \me^{\mi\vec{v}\cdot\vec{\Phi}}\,\alpha^{(n,m,\ell)}(\vec\Phi),
\end{align}
with $\vec\Phi=(\Phi_1,\Phi_2,\Phi_3)$ and $\vec v\cdot\vec\Phi=\sum_{i=1}^3 v_i\Phi_i$. The orthonormality relation reads
\begin{align}
\frac{1}{(2\pi)^3}\int_0^{2\pi} \td^3\Phi \,\me^{\mi (\vec{v}-\vec{v}')\cdot\vec{\Phi}}=\delta_{\vec{v},\vec{v}'}.
\end{align}
For practical implementations, the continuous phase integrals are replaced by discrete Fourier transforms using phase steps $\Phi_i=2\pi k_i/N_\phi$, where $N_\phi$ is the total number of phase steps sampled~\cite{yuenzhou2014ultrafast}. 

It is further useful to note that the components of the cavity field and density matrix follow the Fourier symmetry relations 
 \begin{align}
\left[\alpha^{(n,m,l)}_{\vec{v}}\right]^*=\alpha^{(n,m,l)*}_{-\vec{v}},\quad \left[\rho^{(n,m,l)}_{\vec{v}}\right]^\dagger=\rho^{(n,m,l)}_{-\vec{v}},
 \end{align}
i.e., complex conjugation corresponds to flipping the sign of the phase.

\section{Lindblad terms for pure dephasing}
\label{sec:dephasing}

In Lindblad form, pure dephasing  of a harmonic oscillator can be modeled by a jump operator proportional to $\hat{n}=\hat{b}^\dagger \hat{b}$, yielding the dissipator
\begin{align}
\mathcal{D}_\phi^{(\mathrm{ho})}[\rho]
=
\gamma_\phi \Big(
\hat{n}\rho \hat{n}-\tfrac{1}{2}\{\hat{n}^2,\rho\}
\Big).
\label{eq:lindblad_dephasing_ho}
\end{align}
In the Fock basis, this leads to an exponential decay of the off-diagonal elements,
\begin{align}
\dot\rho_{kl}
=
-\frac{\gamma_\phi}{2}(k-l)^2\,\rho_{kl},
\label{eq:ho_dephasing_rates}
\end{align}
while populations ($k=l$) remain unaffected. In particular, neighboring coherences dephase at the same rate, whereas the $0\!\leftrightarrow\!2$ coherence dephases four times faster than the $0\!\leftrightarrow\!1$ coherence.

Guided by this harmonic-oscillator picture, we implement pure dephasing
in the reduced 3LS by truncating the number operator
to the subspace $\{\ket{g},\ket{e},\ket{f}\}\equiv\{\ket{0},\ket{1},\ket{2}\}$.
The corresponding collapse operator then reads
\begin{equation}
\hat{L}_\phi=
\sqrt{\gamma_\phi}
\left(
\ket{e}\!\bra{e}
+2\,\ket{f}\!\bra{f}
\right),
\label{eq:lindblad_number_3ls}
\end{equation}
and the dissipator is given by
\begin{align}
\label{eq:dephasingdiss}
\mathcal{D}_\phi[\rho]
=
\hat{L}_\phi^{\phantom{\dagger}} \rho \hat{L}_\phi^\dagger
-\tfrac12 \{ \hat{L}_\phi^\dagger \hat{L}_\phi^{\phantom{\dagger}} , \rho \}.
\end{align}

In the plots showing comparison to experimental data, we assign different linewidths to the transitions to more closely reflect the experimental situation.

\section{Perturbative EoM for the anharmonic 3LS}
\label{sec:eoms}

\begin{figure}[h]
    \centering
    \includegraphics[width=0.65\textwidth]{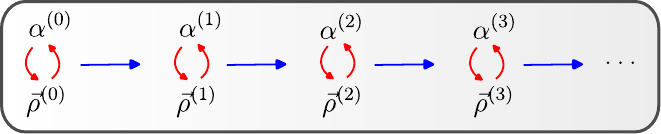}
    \caption{Schematic illustration of the iterative construction of the nonlinear cavity field and density matrix. Starting from the zeroth-order initial states $\alpha^{(0)}$, $\vec{\rho}^{(0)}$, successive coupled light-matter orders are obtained, where each order is sourced by the lower-order contributions.}
    \label{figS2}
\end{figure}

Here, we present the explicit EoM for the cavity field 
and for the components of the molecular density matrix that arise from the perturbative 
amplitude-phase expansion for the anharmonic 3LS. These expressions constitute the basis for our numerical 
implementation and are used to generate all results presented in the main text. The basic principle of the iterative construction of the cavity field and density matrix from lower-order contributions is illustrated in Fig.~\ref{figS2}. We assume initial conditions $\alpha^{(0,0,0)}=\alpha(0)=0$, $\rho^{(0,0,0)}=\rho (0)=\ket{g}\!\bra{g}$.

The EoM are derived and numerically implemented in a rotating frame at the central pulse frequency $\omega_\ell$. The detunings are defined as $\Delta_{ge}=\omega_\ell-\omega_e$, and $\Delta_{ef}=\omega_\ell-(\omega_f-\omega_e)$ and the cavity couplings as $g_{ge}=E_0\mu_{ge}$, $g_{ef}=E_0\mu_{ef}$. The anharmonic shift of the second excited state is given by $\Delta=\Delta_{ef}-\Delta_{ge}$.

\subsection{First order.}

The first order (linear) equations are given by

\begin{subequations}
\begin{align}
\dot{\alpha}^{(1,0,0)}_{\Phi_1}
&=-\left(\frac{\kappa}{2}-\mi\Delta_c\right)\alpha^{(1,0,0)}_{\Phi_1}
-\mi \mathcal{N}g_{ge}\rho_{eg, {\Phi_1}}^{(1,0,0)}
-f_1 (t-t_1),\\
\dot{\alpha}^{(0,1,0)}_{\Phi_2}
&=-\left(\frac{\kappa}{2}-\mi\Delta_c\right)\alpha^{(0,1,0)}_{\Phi_2}
-\mi \mathcal{N}g_{ge}\rho_{eg, {\Phi_2}}^{(0,1,0)}
-f_2 (t-t_2),\\
\dot{\alpha}^{(0,0,1)}_{\Phi_3}
&=-\left(\frac{\kappa}{2}-\mi\Delta_c\right)\alpha^{(0,0,1)}_{\Phi_3}
-\mi \mathcal{N}g_{ge}\rho_{eg, {\Phi_3}}^{(0,0,1)}
-f_3 (t-t_3),
\\[1em]
\dot{\rho}_{eg, {\Phi_1}}^{(1,0,0)}
&=-\left(\frac{\gamma_\phi}{2}-\mi\Delta_{ge}\right)
\rho_{eg, {\Phi_1}}^{(1,0,0)}
-\mi g_{ge}\alpha^{(1,0,0)}_{\Phi_1},\\
\dot{\rho}_{eg, {\Phi_2}}^{(0,1,0)}
&=-\left(\frac{\gamma_\phi}{2}-\mi\Delta_{ge}\right)
\rho_{eg, {\Phi_2}}^{(0,1,0)}
-\mi g_{ge}\alpha^{(0,1,0)}_{\Phi_2},\\
\dot{\rho}_{eg, {\Phi_3}}^{(0,0,1)}
&=-\left(\frac{\gamma_\phi}{2}-\mi\Delta_{ge}\right)
\rho_{eg, {\Phi_3}}^{(0,0,1)}
-\mi g_{ge}\alpha^{(0,0,1)}_{\Phi_3}.
\end{align}
\end{subequations}
Thus, at first order, each pulse independently generates a coherent cavity field as well as a coherence between $\ket{g}$ and $\ket{e}$, which directly inherit the phase of the corresponding incoming pulse.

\subsection{Second order.}

The phase components of the second-order populations and coherences are given by 
\begin{subequations}
\label{eq:secondorder}
\begin{align}
\dot{\rho}_{ee, \Phi_1-\Phi_2}^{(1,1,0)}
&=-\mi g_{ge}\alpha^{(1,0,0)}_{\Phi_1}\rho_{ge,-\Phi_2}^{(0,1,0)}
+\mi g_{ge}\alpha^{(0,1,0)*}_{-\Phi_2}\rho^{(1,0,0)}_{eg, \Phi_1},\\
\dot{\rho}_{ee, -\Phi_1+\Phi_2}^{(1,1,0)}
&=-\mi g_{ge}\alpha^{(0,1,0)}_{\Phi_2}\rho_{ge,-\Phi_1}^{(0,1,0)}
+\mi g_{ge}\alpha^{(1,0,0)*}_{-\Phi_1}\rho^{(0,1,0)}_{eg, \Phi_2},\\
\dot{\rho}_{ee, \Phi_1-\Phi_3}^{(1,0,1)}
&=-\mi g_{ge}\alpha^{(1,0,0)}_{\Phi_1}\rho_{ge,-\Phi_3}^{(0,0,1)}
+\mi g_{ge}\alpha^{(0,0,1)*}_{-\Phi_3}\rho^{(1,0,0)}_{eg, \Phi_1},\\
\dot{\rho}_{ee, -\Phi_1+\Phi_3}^{(1,0,1)}
&=-\mi g_{ge}\alpha^{(0,0,1)}_{\Phi_3}\rho_{ge,-\Phi_1}^{(1,0,0)}
+\mi g_{ge}\alpha^{(1,0,0)*}_{-\Phi_1}\rho^{(0,0,1)}_{eg, \Phi_3},\\
\dot{\rho}_{ee, \Phi_2-\Phi_3}^{(0,1,1)}
&=-\mi g_{ge}\alpha^{(0,1,0)}_{\Phi_2}\rho_{ge,-\Phi_3}^{(0,0,1)}
+\mi g_{ge}\alpha^{(0,0,1)*}_{-\Phi_3}\rho^{(0,1,0)}_{eg, \Phi_2},\\
\dot{\rho}_{ee, -\Phi_2+\Phi_3}^{(0,1,1)}
&=-\mi g_{ge}\alpha^{(0,0,1)}_{\Phi_3}\rho_{ge,-\Phi_2}^{(0,1,0)}
+\mi g_{ge}\alpha^{(0,1,0)*}_{-\Phi_2}\rho^{(0,0,1)}_{eg, \Phi_3},\\
\dot{\rho}_{fg, \Phi_1+\Phi_2}^{(1,1,0)}
&=-\left[2\gamma_\phi-\mi(\Delta_{ef}+\Delta_{ge})\right]\rho_{fg, \Phi_1+\Phi_2}^{(1,1,0)}
-\mi g_{ef}\alpha^{(1,0,0)}_{\Phi_1}\rho^{(0,1,0)}_{eg, \Phi_2}
-\mi g_{ef}\alpha^{(0,1,0)}_{\Phi_2}\rho^{(1,0,0)}_{eg, \Phi_1},\\
\dot{\rho}_{fg, \Phi_1+\Phi_3}^{(1,0,1)}
&=-\left[2\gamma_\phi-\mi(\Delta_{ef}+\Delta_{ge})\right]\rho_{fg, \Phi_1+\Phi_3}^{(1,0,1)}
-\mi g_{ef}\alpha^{(1,0,0)}_{\Phi_1}\rho^{(0,0,1)}_{eg, \Phi_3}
-\mi g_{ef}\alpha^{(0,0,1)}_{\Phi_3}\rho^{(1,0,0)}_{eg, \Phi_1},\\
\dot{\rho}_{fg, \Phi_2+\Phi_3}^{(0,1,1)}
&=-\left[2\gamma_\phi-\mi(\Delta_{ef}+\Delta_{ge})\right]\rho_{fg, \Phi_2+\Phi_3}^{(0,1,1)}
-\mi g_{ef}\alpha^{(0,1,0)}_{\Phi_2}\rho^{(0,0,1)}_{eg, \Phi_3}
-\mi g_{ef}\alpha^{(0,0,1)}_{\Phi_3}\rho^{(0,1,0)}_{eg, \Phi_2}.
\end{align}
\end{subequations}
Hence, the second-order response comprises six independent phase combinations leading to population in the intermediate state $\ket{e}$, and three phase combinations giving rise to coherences between $\ket{f}$ and $\ket{g}$. No cavity field is generated at second order.

\subsection{Third order.}

The matter phase contributions at third order are given by
\begin{subequations}
\label{eq:thirdorderphase}
\begin{align}
\dot{\rho}_{eg, \Phi_1+\Phi_2-\Phi_3}^{(1,1,1)}
&=-(\gamma_\phi/2-\mi\Delta_{ge})\rho_{eg, \Phi_1+\Phi_2-\Phi_3}^{(1,1,1)}
-\mi g_{ge}\alpha^{(1,1,1)}_{\Phi_1+\Phi_2-\Phi_3}
+2\mi g_{ge}\alpha^{(1,0,0)}_{\Phi_1}\rho_{ee, \Phi_2-\Phi_3}^{(0,1,1)}
\nonumber\\
&\quad
+2\mi g_{ge}\alpha^{(0,1,0)}_{\Phi_2}\rho_{ee, \Phi_1-\Phi_3}^{(1,0,1)}
-\mi g_{ef}\alpha^{(0,0,1)*}_{-\Phi_3}\rho_{fg, \Phi_1+\Phi_2}^{(1,1,0)},
\label{eq:thirdorderphase_a}\\[6pt]
\dot{\rho}_{eg, \Phi_1-\Phi_2+\Phi_3}^{(1,1,1)}
&=-(\gamma_\phi/2-\mi\Delta_{ge})\rho_{eg, \Phi_1-\Phi_2+\Phi_3}^{(1,1,1)}
-\mi g_{ge}\alpha^{(1,1,1)}_{\Phi_1-\Phi_2+\Phi_3}
+2\mi g_{ge}\alpha^{(1,0,0)}_{\Phi_1}\rho_{ee, -\Phi_2+\Phi_3}^{(0,1,1)}
\nonumber\\
&\quad
+2\mi g_{ge}\alpha^{(0,0,1)}_{\Phi_3}\rho_{ee, \Phi_1-\Phi_2}^{(1,1,0)}
-\mi g_{ef}\alpha^{(0,1,0)*}_{-\Phi_2}\rho_{fg, \Phi_1+\Phi_3}^{(1,0,1)}
\label{eq:thirdorderphase_b},\\[6pt]
\dot{\rho}_{eg, -\Phi_1+\Phi_2+\Phi_3}^{(1,1,1)}
&=-(\gamma_\phi/2-\mi\Delta_{ge})\rho_{eg, -\Phi_1+\Phi_2+\Phi_3}^{(1,1,1)}
-\mi g_{ge}\alpha^{(1,1,1)}_{-\Phi_1+\Phi_2+\Phi_3}
+2\mi g_{ge}\alpha^{(0,1,0)}_{\Phi_2}\rho_{ee, -\Phi_1+\Phi_3}^{(1,0,1)}
\nonumber\\
&\quad
+2\mi g_{ge}\alpha^{(0,0,1)}_{\Phi_3}\rho_{ee, -\Phi_1+\Phi_2}^{(1,1,0)}
-\mi g_{ef}\alpha^{(1,0,0)*}_{-\Phi_1}\rho_{fg, \Phi_2+\Phi_3}^{(0,1,1)},
\label{eq:thirdorderphase_c}\\[6pt]
\dot{\rho}_{fe, \Phi_1+\Phi_2-\Phi_3}^{(1,1,1)}
&=-\left(\gamma_\phi/2-\mi\Delta_{ef}\right)\rho^{(1,1,1)}_{fe, \Phi_1+\Phi_2-\Phi_3}
-\mi g_{ef}\alpha^{(0,1,0)}_{\Phi_2}\rho_{ee, \Phi_1-\Phi_3}^{(1,0,1)}
\nonumber\\
&\quad
-\mi g_{ef}\alpha^{(1,0,0)}_{\Phi_1}\rho_{ee, \Phi_2-\Phi_3}^{(0,1,1)}
+\mi g_{ge}\alpha^{(0,0,1)*}_{-\Phi_3}\rho_{fg, \Phi_1+\Phi_2}^{(1,1,0)},
\label{eq:thirdorderphase_d}\\[6pt]
\dot{\rho}_{fe, \Phi_1-\Phi_2+\Phi_3}^{(1,1,1)}
&=-\left(\gamma_\phi/2-\mi\Delta_{ef}\right)\rho^{(1,1,1)}_{fe, \Phi_1-\Phi_2+\Phi_3}
-\mi g_{ef}\alpha^{(0,0,1)}_{\Phi_3}\rho_{ee, \Phi_1-\Phi_2}^{(1,1,0)}
\nonumber\\
&\quad
-\mi g_{ef}\alpha^{(1,0,0)}_{\Phi_1}\rho_{ee, -\Phi_2+\Phi_3}^{(0,1,1)}
+\mi g_{ge}\alpha^{(0,1,0)*}_{-\Phi_2}\rho_{fg, \Phi_1+\Phi_3}^{(1,0,1)},
\label{eq:thirdorderphase_e}\\[6pt]
\dot{\rho}_{fe, -\Phi_1+\Phi_2+\Phi_3}^{(1,1,1)}
&=-\left(\gamma_\phi/2-\mi\Delta_{ef}\right)\rho^{(1,1,1)}_{fe, -\Phi_1+\Phi_2+\Phi_3}
-\mi g_{ef}\alpha^{(0,0,1)}_{\Phi_3}\rho_{ee, -\Phi_1+\Phi_2}^{(1,1,0)}
\nonumber\\
&\quad
-\mi g_{ef}\alpha^{(0,1,0)}_{\Phi_2}\rho_{ee, -\Phi_1+\Phi_3}^{(1,0,1)}
+\mi g_{ge}\alpha^{(1,0,0)*}_{-\Phi_1}\rho_{fg, \Phi_2+\Phi_3}^{(0,1,1)},
\label{eq:thirdorderphase_f}
\end{align}
\end{subequations}
Therefore, possible phase combinations are $\Phi_1+\Phi_2-\Phi_3$, $\Phi_1-\Phi_2+\Phi_3$, and $-\Phi_1+\Phi_2+\Phi_3$ (and the complex conjugates). The field components driven at third order are therefore given by
\begin{subequations}
\begin{align}
\dot{\alpha}^{(1,1,1)}_{\Phi_1+\Phi_2-\Phi_3}
&=-\left(\kappa/2-\mi\Delta_c\right)\alpha^{(1,1,1)}_{\Phi_1+\Phi_2-\Phi_3}
-\mi g_{ge}\mathcal{N}\rho_{eg, \Phi_1+\Phi_2-\Phi_3}^{(1,1,1)}
-\mi g_{ef}\mathcal{N}\rho_{fe, \Phi_1+\Phi_2-\Phi_3}^{(1,1,1)},\\
\dot{\alpha}^{(1,1,1)}_{\Phi_1-\Phi_2+\Phi_3}
&=-\left(\kappa/2-\mi\Delta_c\right)\alpha^{(1,1,1)}_{\Phi_1-\Phi_2+\Phi_3}
-\mi g_{ge}\mathcal{N}\rho_{eg, \Phi_1-\Phi_2+\Phi_3}^{(1,1,1)}
-\mi g_{ef}\mathcal{N}\rho_{fe, \Phi_1-\Phi_2+\Phi_3}^{(1,1,1)},\\
\dot{\alpha}^{(1,1,1)}_{-\Phi_1+\Phi_2+\Phi_3}
&=-\left(\kappa/2-\mi\Delta_c\right)\alpha^{(1,1,1)}_{-\Phi_1+\Phi_2+\Phi_3}
-\mi g_{ge}\mathcal{N}\rho_{eg, -\Phi_1+\Phi_2+\Phi_3}^{(1,1,1)}
-\mi g_{ef}\mathcal{N}\rho_{fe, -\Phi_1+\Phi_2+\Phi_3}^{(1,1,1)}.
\end{align}
\end{subequations}
To facilitate comparison with the experiment, we allow the $e\leftrightarrow g$ and $f\leftrightarrow\!e$ coherences to have different linewidths, replacing $\gamma_\phi$ by $\gamma_\phi^{ge}$ and $\gamma_\phi^{ef}$, respectively. For the $g\leftrightarrow f$ coherence, we assume a decay rate given by $\gamma_\phi^{ge}+\gamma_\phi^{ef}$. This choice ensures that, in the limit $\gamma_\phi^{ge}=\gamma_\phi^{ef}$, the model reduces to the usual harmonic-dephasing result.

\begin{figure}[h]
    \centering
    \includegraphics[width=1.0\textwidth]{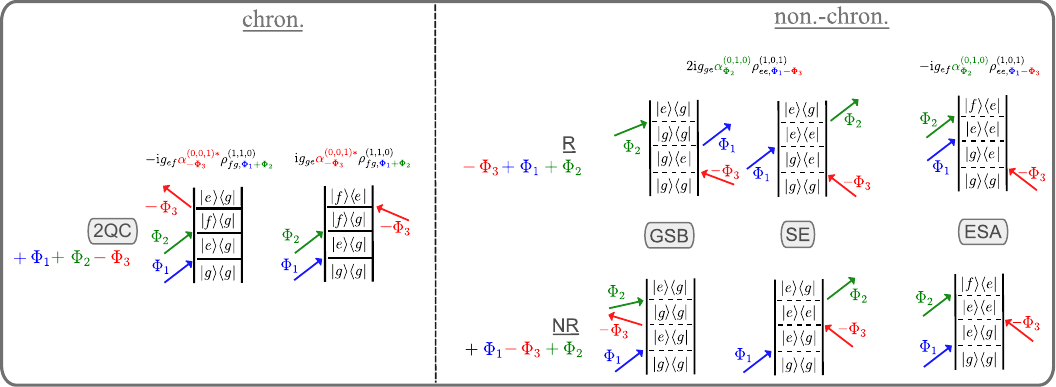}
    \caption{Sketch of chronological 2QC pathways (left) as well as examples of non-chronological ground state bleach (GSB), stimulated emission (SE) and excited state absorption (ESA) pathways contributing to the 2QC component of the third-order molecular density matrix. Additional non-chronological contributions arise from terms $2\mi g_{ge}\alpha_{\Phi_1}^{(1,0,0)}\rho_{ee,\Phi_2-\Phi_3}^{(0,1,1)}$ and $-\mi g_{ef}\alpha_{\Phi_1}^{(1,0,0)}\rho_{ee,\Phi_2-\Phi_3}^{(0,1,1)}$ which correspond to exchanging $\Phi_1$ and $\Phi_2$ in the above. For brevity, we indicate only the phases which label the cavity fields acting on the molecular density matrix.}
    \label{figS1}
\end{figure}

\subsection{Feynman diagrams}

From the EoM in~\eqref{eq:thirdorderphase}, the corresponding double-sided Feynman diagrams can be directly constructed by tracking the sequence of light-matter interactions and their associated phase combinations. In Fig.~\ref{figS1}, we show the diagrams for the  2QC pathways of the molecular density matrix associated specifically with Eqs.~\eqref{eq:thirdorderphase_a} and \eqref{eq:thirdorderphase_d}, which generate the $\Phi_1+\Phi_2-\Phi_3$ phase-matching condition. Both chronological and non-chronological contributions arise, reflecting the effective scrambling of the pulses inside the cavity.

\section{Definition of differential transmission}

Following Ref.~\cite{reitz2025nonlinear}, we define the DT signal for each phase component as the difference in transmission between all 3 pulses on and only the last pulse (probe) on:
\begin{align}
\label{eq:difft}
\nonumber\Delta \mathcal{T}_{\vec{v}\cdot\vec{\Phi}} (\tau, T, \omega_3)&=\mathcal{T}_{\vec{v}\cdot\vec{\Phi}} ^\mathrm{pump-on}(\tau, T, \omega_3)-\mathcal{T}_{\vec{v}\cdot\vec{\Phi}} ^\mathrm{pump-off}(\omega_3)\\
\nonumber&\approx\left(\frac{\kappa}{2}\right)^2\left[\frac{|\alpha_{\Phi_3}^{(0,0,1)}(\omega_3)|^2}{f_3(\omega_3)^2}-\frac{|\alpha^{(0,0,1)}_{\Phi_3}(\omega_3)+\eta_1\eta_2\alpha^{(1,1,1)}_{\vec{v}\cdot\vec{\Phi}}(\tau, T,\omega_3)|^2}{f_3(\omega_3)^2}\right]\\
&\approx \left(\frac{\kappa}{2}\right)^2\eta_1\eta_2\frac{2\mathrm{Re}[\alpha_{-\Phi_3}^{(0,0,1)*}(\omega_3)\alpha_{\vec{v}\cdot\vec{\Phi}}^{(1,1,1)}(\tau, T, \omega_3)]}{f_3 (\omega_3)^2}.
\end{align}
In the last step, we retained only the four-wave mixing contribution, neglecting all higher-order nonlinear terms.

\section{Cancellation of nonlinear response in harmonic limit}
\label{sec:harmonic}

To make explicit that our model correctly reproduces the absence of any nonlinear response for a purely harmonic oscillator, let us briefly analyze the harmonic limit of the third-order solution. For simplicity, we consider only a single input pulse $f(t)$, such that $\alpha^{(1,1,1)} \equiv \alpha^{(3)}$, $\rho_{ee}^{(1,1,0)}\equiv\rho^{(2)}$, and analogously for other quantities. We further drop the explicit phase indices for notational clarity, however this can be straightforwardly generalized to the full phase-resolved expressions.

The simplified third-order EoM for a single pulse read in frequency space
\begin{subequations}
\begin{align}
-\mi \omega \alpha^{(3)}(\omega)
&=
-\left( \frac{\kappa}{2} - \mi \Delta_c \right)\alpha^{(3)}(\omega)
-\mi g_{ge}\mathcal{N}\,\rho_{eg}^{(3)}(\omega)
-\mi g_{ef}\mathcal{N}\,\rho_{fe}^{(3)}(\omega),
\\[4pt]
-\mi \omega \rho_{eg}^{(3)}(\omega)
&=
-\left( \frac{\gamma_\phi}{2} - \mi \Delta_{ge} \right)\rho_{eg}^{(3)}(\omega)
-\mi g_{ge}\alpha^{(3)}(\omega)
+2\mi g_{ge}\left[\alpha^{(1)} \ast \rho_{ee}^{(2)}\right](\omega)
-\mi g_{ef}\left[\alpha^{(1)\ast} \ast \rho_{fg}^{(2)}\right](\omega),
\\[4pt]
-\mi \omega \rho_{fe}^{(3)}(\omega)
&=
-\left( \frac{\gamma_\phi}{2} - \mi \Delta_{ef} \right)\rho_{fe}^{(3)}(\omega)
-\mi g_{ef}\left[\alpha^{(1)} \ast \rho_{ee}^{(2)}\right](\omega)
+\mi g_{ge}\left[\alpha^{(1)\ast} \ast \rho_{fg}^{(2)}\right](\omega),
\end{align}
\end{subequations}
where $\ast$ between two quantities denotes convolution. This can be solved for the cavity field
\begin{align}
\label{eq:contributions}
\alpha^{(3)}(\omega)=&\frac{-\alpha^{(1)}(\omega)}{\tilde{f}(\omega)}\mathcal{N}\Bigg[
\underbrace{
+2g_{ge}^2\frac{\left[\alpha^{(1)}\ast\rho_{ee}^{(2)}\right](\omega)}{\frac{\gamma_\phi}{2}-\mi(\omega+\Delta_{ge})}
}_{\text{GSB+SE}}
\;\
\underbrace{
-g_{ef}^2\frac{\left[\alpha^{(1)}\ast\rho_{ee}^{(2)}\right] (\omega)}{\frac{\gamma_\phi}{2}-\mi(\omega+\Delta_{ef})}
}_{\text{ESA}}\\ \nonumber
&\hspace{3.2cm}
\underbrace{
-g_{ge}g_{ef}\left[\alpha^{(1)*}\ast\rho_{fg}^{(2)}\right](\omega)
\left(\frac{1}{\frac{\gamma_\phi}{2}-\mi(\omega+\Delta_{ge})}
-\frac{1}{\frac{\gamma_\phi}{2}-\mi(\omega+\Delta_{ef})}\right)
}_{\text{2QC}}
\Bigg].
\end{align}
\begin{figure}[b]
    \centering
    \includegraphics[width=0.95\textwidth]{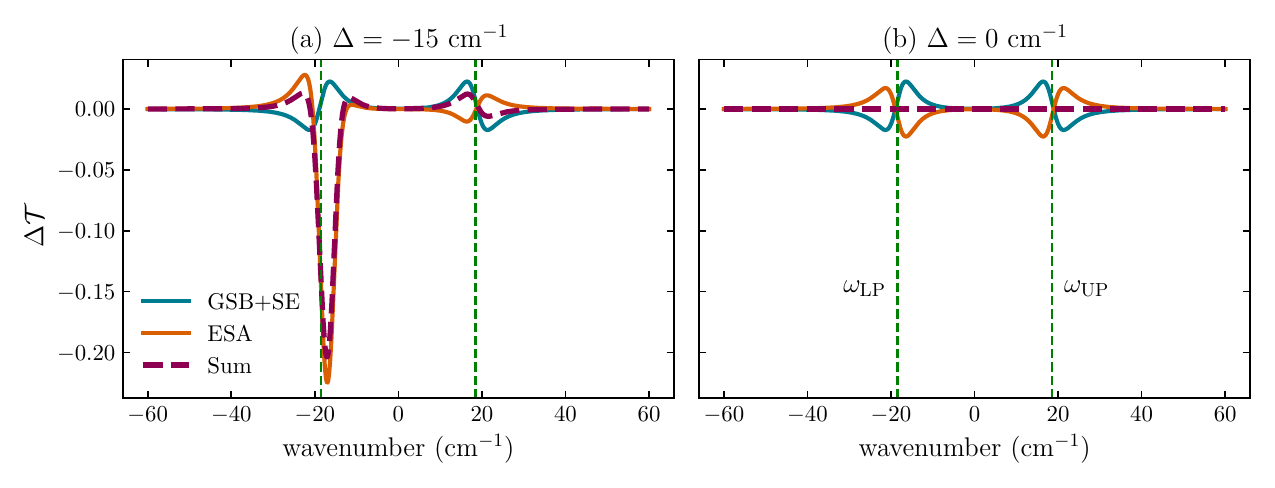}
    \caption{Individual contributions [as described by Eq.~\eqref{eq:contributions}] to the differential transmission 
    $\Delta\mathcal{T}(\omega)\sim\mathrm{Re}[\alpha^{(1)*}(\omega)\alpha^{(3)}(\omega)]$ in the stationary limit arising from GSB+SE and ESA (solid curves), as well as their sum (dashed curve). Panel (a) shows the case with finite anharmonicity, $\Delta=-15~\mathrm{cm}^{-1}$, while panel (b) corresponds to the harmonic limit, $\Delta=0$. 
    Parameters used are $g_{ge}\sqrt{\mathcal{N}}=18.5~\mathrm{cm}^{-1}$, 
    $g_{ef}=\sqrt{2}\,g_{ge}$, 
    $\kappa=11~\mathrm{cm}^{-1}$, 
    $\gamma_\phi=6~\mathrm{cm}^{-1}$ (harm.~dephasing), $\delta=0$, $\omega_\ell=\omega_c=\omega_e$, and the results are plotted in the rotating frame. 
    The dashed green vertical lines indicate the linear polariton frequencies at $\omega_{\mathrm{UP}/\mathrm{LP}}=\pm g_{ge}\sqrt{\mathcal{N}}$.}
    \label{figS4}
\end{figure}
The second order quantities can again be expressed in terms of linear quantities
\begin{subequations}
\begin{align}
\rho_{ee}^{(2)}(\omega)
&=\frac{2\mi g_{ge}}{\omega}
\mathrm{Im}\left[\alpha^{(1)}\ast\rho_{ge}^{(1)}\right](\omega),\\
\rho_{fg}^{(2)}(\omega)
&=\frac{-\mi g_{ef}\left[\alpha^{(1)}\ast\rho_{eg}^{(1)}\right](\omega)}
{2\gamma_\phi-\mi(\omega+\Delta_{ef}+\Delta_{ge})},
\end{align}
\end{subequations}
and the linear (first-order) quantities are given by
\begin{subequations}
\begin{align}
 \alpha^{(1)}(\omega)
&=\frac{-\tilde{f}(\omega)}
{\frac{\kappa}{2}-\mi(\omega+\Delta_c)
+\frac{\mathcal{N}g_{ge}^2}{\frac{\gamma_\phi}{2}-\mi(\omega+\Delta_{ge})}},\\
\rho_{eg}^{(1)}(\omega)
&=\frac{-\mi g_{ge}\alpha^{(1)}(\omega)}
{\frac{\gamma_\phi}{2}-\mi(\omega+\Delta_{ge})}.
\end{align}
\end{subequations}
In Eq.~\eqref{eq:contributions} we have identified the individual GSB, SE, ESA, and 2QC contributions to the third-order cavity field. Here, GSB and SE give identical contributions and therefore appear as a single term with an overall factor of 2. Both, as well as ESA, require population in $\ket{e}$, while the 2QC term requires coherence between $\ket{g}$ and $\ket{f}$ at second order. Importantly, the third-order response retains the familiar decomposition into GSB+SE, ESA, and 2QC pathways known from bare-molecule nonlinear spectroscopy. The difference is that, in the cavity, these pathways are driven by the linear cavity field $\alpha^{(1)}$, not directly by the laser pulses, and the resulting third-order field is further filtered by the linear polaritonic response window encoded in $\alpha^{(1)}(\omega)$.

In the harmonic limit, defined by $\Delta_{ge}=\Delta_{ef}$ and 
$g_{ef}=\sqrt{2}\,g_{ge}$, all third-order pathways in Eq.~\eqref{eq:contributions} cancel exactly, i.e., $\alpha^{(3)}=0$. 
The two 2QC contributions cancel each other, while the ESA pathway exactly cancels the combined GSB/SE contributions due to the oscillator-like scaling of the transition strengths. Fig.~\ref{figS4} shows a plot of the GSB, SE and ESA contributions of $\alpha^{(3)}(\omega)$ to $\Delta\mathcal{T}(\omega)$ in the stationary limit, i.e., assuming constant population $\rho_{ee}(\omega)=\rho_{ee}$, in the anharmonic case as well as in the harmonic limit.

\begin{figure}[t]
    \centering
    \includegraphics[width=1.0\textwidth]{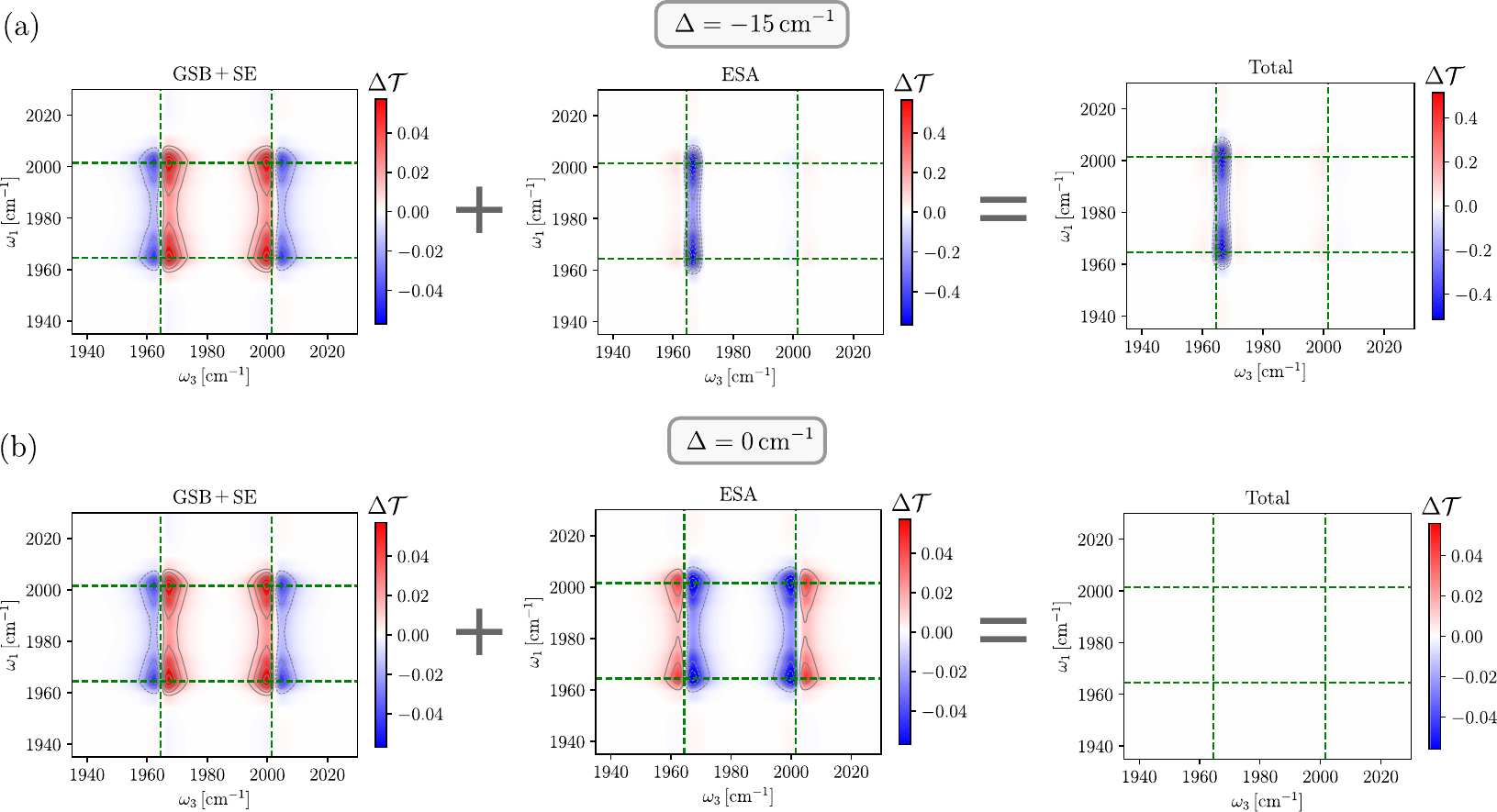}
    \caption{GSB+SE (left column) and ESA (middle column) contributions as well as sum (right column) of 1Q 2D spectra (NR+R) for mechanical anharmonicities (a) $\Delta=-15~\mathrm{cm}^{-1}$ and (b) $\Delta=0~\mathrm{cm}^{-1}$ (harm.~limit), at a waiting time of $T=18~\kappa^{-1}$. Other parameters are $g_{ge}\sqrt{\mathcal{N}}=18.5~\mathrm{cm}^{-1}$, 
    $g_{ef}=\sqrt{2}\,g_{ge}$, 
    $\kappa=11~\mathrm{cm}^{-1}$, 
    $\gamma_\phi=6~\mathrm{cm}^{-1}$ (harm.~dephasing), $\delta=0$, $ \omega_\ell=\omega_c=\omega_e$. 
    The dashed green lines indicate the linear polariton frequencies $\omega_{\mathrm{UP}/\mathrm{LP}}$ along both axes. }
    \label{figS5}
\end{figure}

\section{GSB, SE and ESA contributions to 2D spectra}
\label{sec:contributions}

Fig.~\ref{figS5} shows the contributions of GSB+SE (combined) and ESA to the 2D 1Q spectra, for the sum of NR+R signals. We present results for both the anharmonic case $\Delta=-15~\mathrm{cm}^{-1}$ and the harmonic limit $\Delta=0$. These contributions are obtained by taking the third-order EoM for the density matrix \eqref{eq:thirdorderphase} and only keeping terms describing a specific process (GSB and SE give identical contribution for the system considered here and are therefore taken together; also see Sec.~\ref{sec:harmonic} above). The GSB+SE contribution is identical for any degree of anharmonicity, as it does not involve the doubly excited state. Note that, due to the overlap of the fields within the cavity (i.e., the third pulse can act before the second pulse), the NR signal also contains a small but non-vanishing 2QC contribution. This contribution becomes negligible in the limit of large waiting times, $T \gg \kappa^{-1}$. Therefore, at long waiting times as shown here, the total signal is well-approximated by the sum of the contributions in Fig.~\ref{figS5}, which cancel in the harmonic limit $\Delta=0$.






\section{Excitation-induced dephasing (EID) \& polariton bleach}
\label{sec:polaritonbleach}

To reproduce the polariton bleach observed at short times, we include excitation-induced dephasing (EID) by introducing a population-dependent dephasing rate in the dissipator of Eq.~\eqref{eq:dephasingdiss}. Expanding the dephasing rate in a Taylor series about vanishing excited-state population and retaining only the first-order correction, we can approximate~\cite{wang1993transient}
\begin{align}
\gamma_\phi(\rho_{ee})
\approx
\gamma_\phi(0)
+
\left.\frac{\partial \gamma_\phi}{\partial \rho_{ee}}\right|_{\rho_{ee}=0}\rho_{ee}
\equiv
\gamma_{\phi,0}+\beta\,\rho_{ee},
\end{align}
where
\(
\beta=\gamma_\phi'
\)
denotes the slope of the population-dependent dephasing rate, and thus the strength of the EID, while
\(
\gamma_{\phi,0}\equiv \gamma_\phi(0)
\)
denotes the bare dephasing rate in the absence of excited-state population.

Within the perturbative expansion, assuming initial condition $\rho_{ee}(0)=\ket{g}\bra{g}$, the excited-state population first appears at second order, \(\rho_{ee}^{(2)}\), so that the EID contribution enters the equation of motion for the \(e\leftrightarrow g\) coherence at third order through the product \(\rho_{ee}^{(2)}\rho_{eg}^{(1)}\). Consequently, the \(e \leftrightarrow g\) coherence acquires an additional third-order damping term of the form
\begin{align}
\dot{\rho}_{eg}^{(3)} = -\left(\frac{\gamma_{\phi,0}}{2}-\mi\Delta_{ge}\right)
\rho_{eg}^{(3)}
+\hdots -\frac{\beta}{2}\rho_{ee}^{(2)}\rho_{eg}^{(1)},
\end{align}
which describes an enhanced decay of the third-order \(eg\) coherence in the presence of excited-state population. By contrast, no analogous contribution arises for the \(f\leftrightarrow e\) coherence at this order, since there is no first-order \(f\leftrightarrow e\) coherence with which the second-order population could combine. Hence, within this perturbative framework, EID selectively enhances the decay of the third-order \(e\leftrightarrow g\) transition, while leaving the third-order \(f\leftrightarrow e\) coherence unaffected to leading order.

For the three-pulse sequence considered here, the EID contribution enters the third-order \(e\leftrightarrow g\) pathways through the population-coherence products whose phases sum to the corresponding phase combinations. The EoM for the R and NR pathways in Eq.~\eqref{eq:thirdorderphase} can then be generalized to 
\begin{subequations}
\begin{align}
\dot{\rho}_{eg,\Phi_1-\Phi_2+\Phi_3}^{(1,1,1)}
&=
-\left(\frac{\gamma_{\phi,0}}{2}-\mi\Delta_{ge}\right)
\rho_{eg,\Phi_1-\Phi_2+\Phi_3}^{(1,1,1)}
+\hdots
-\frac{\beta}{2}\rho_{eg,\Phi_1}^{(1,0,0)}\rho_{ee,-\Phi_2+\Phi_3}^{(0,1,1)}
-\frac{\beta}{2}\rho_{eg,\Phi_3}^{(0,0,1)}\rho_{ee,\Phi_1-\Phi_2}^{(1,1,0)},
\\[6pt]
\dot{\rho}_{eg,-\Phi_1+\Phi_2+\Phi_3}^{(1,1,1)}
&=
-\left(\frac{\gamma_{\phi,0}}{2}-\mi\Delta_{ge}\right)
\rho_{eg,-\Phi_1+\Phi_2+\Phi_3}^{(1,1,1)}
+\hdots
-\frac{\beta}{2}\rho_{eg,\Phi_3}^{(0,0,1)}\rho_{ee,-\Phi_1+\Phi_2}^{(1,1,0)}
-\frac{\beta}{2}\rho_{eg,\Phi_2}^{(0,1,0)}\rho_{ee,-\Phi_1+\Phi_3}^{(1,0,1)}.
\end{align}
\end{subequations}

\begin{figure}[b]
    \centering
    \includegraphics[width=1.0\linewidth]{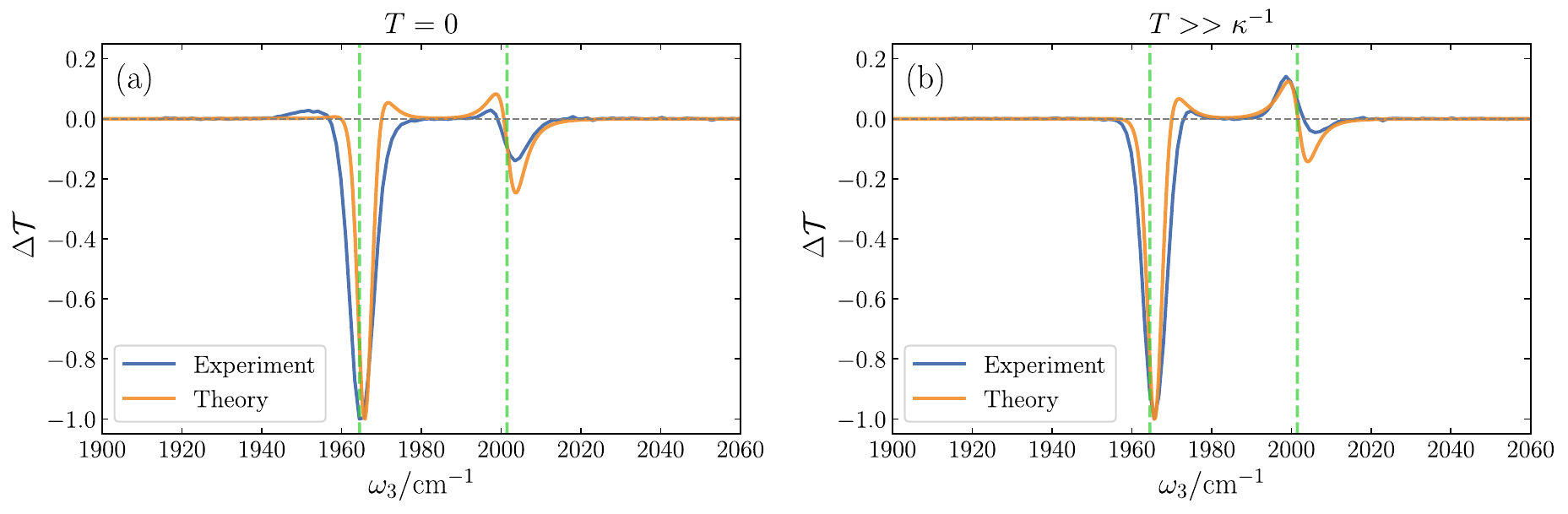}
    \caption{Comparison between experimental (blue) and theoretical (orange) 1D pump-probe spectra (R+NR, normalized to maximum absolute value) at (a) zero delay, $T=0$, and (b) long delay times, $T\gg\kappa^{-1}$, including EID with $\beta=10.0~\mathrm{cm}^{-1}$. All other parameters are identical to Table~\ref{tab:exp_params} in the main text. The dashed green vertical lines indicate the linear polariton frequencies $\omega_{\mathrm{UP}/\mathrm{LP}}$.}
    \label{fig:figS6}
\end{figure}

Fig.~\ref{fig:figS6} shows a comparison between experiment and theory for the 1D pump-probe spectra when EID is included. We find that introducing $\beta$ qualitatively reproduces the polariton bleach observed at short waiting times, while still retaining the spectral contraction around $\omega_{\mathrm{UP}}$ at long waiting times $T\gg\kappa^{-1}$.\\

\begin{figure}[t]
    \centering
    \includegraphics[width=1.0\linewidth]{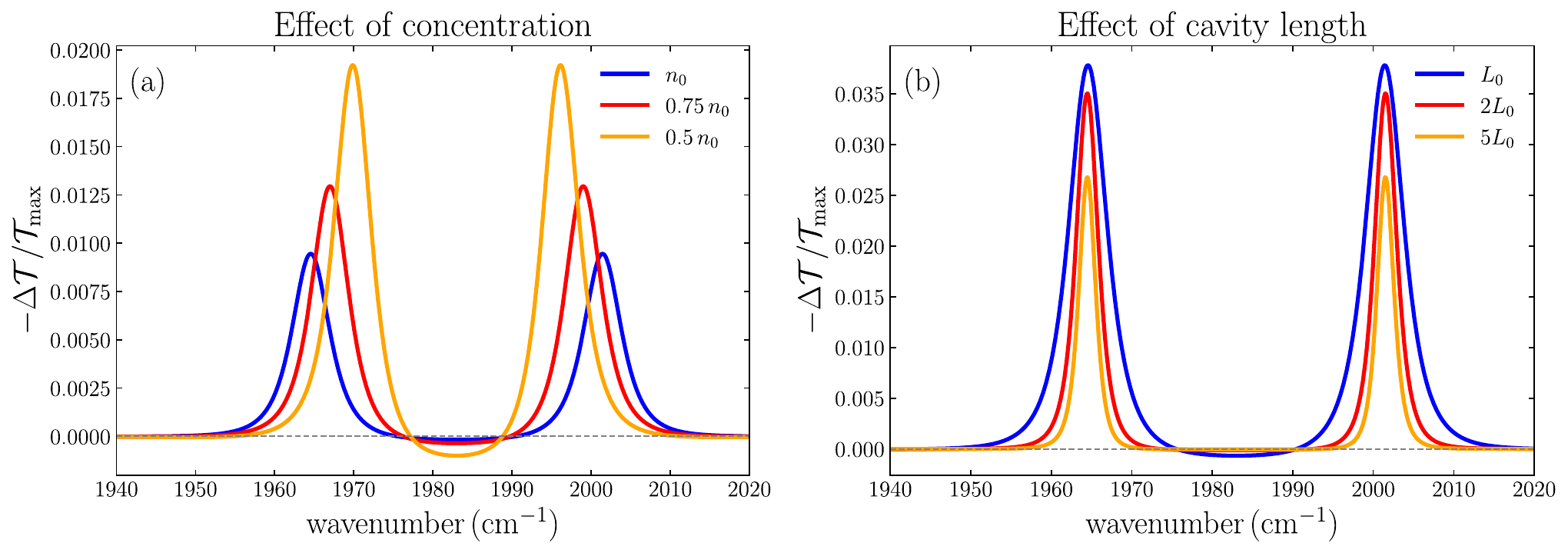}
    \caption{1D pump-probe spectra (R+NR) at zero delay $T=0$ for (a) varying concentration $n$ and (b) varying cavity length $L$ (or equivalently, varying $\kappa$), normalized to the maximum transmission $\mathcal{T}_\mathrm{max}$ of the linear signal of each curve, as in Ref.~\cite{xiang2019manipulating}. To isolate the effect of the polariton bleach signal, we set $g_{ef}=0$ and neglect the terms that give rise to derivative lineshapes, i.e., contraction. The reference values $n_0$ and $L_0$ (or $\kappa_0$) correspond to those chosen in all other plots; all other parameters are also kept the same.}
    \label{fig:figS7}
\end{figure}

Further, in Ref.~\cite{xiang2019manipulating} (Figs.~1(c) and 2), the dependence of the bleach signal on cavity length $L$ and molecular concentration $n=\mathcal{N}/\mathcal{V}$ was investigated, and a decrease of the bleach was observed with increasing concentration as well as with increasing cavity length. In Fig.~\ref{fig:figS7}, we analyze the dependence of the bleach  on these two quantities within our model based on EID. To isolate the pure absorptive bleach contribution on $\Delta\mathcal{T}$, we neglect ESA by setting $g_{ef}=0$, as well as the terms responsible for the contraction effect yielding derivative lineshapes, which are of the form $2\mathrm{i}g_{ge}\alpha^{(1)}\rho_{ee}^{(2)}$. The Figure shows that the model reproduces the experimentally observed trends. In the case of concentration, decreasing $n$ reduces the collective light-matter coupling and hence the Rabi splitting, since $g_{ge}\sqrt{\mathcal{N}}\propto n$. This brings the polariton resonances into greater overlap with the molecular absorption window, resulting in a larger molecular excited-state population $\rho_{ee}$ and therefore a stronger bleach signal. The dependence on cavity length arises primarily through the cavity decay rate, since for a coplanar cavity $\kappa (L) \propto 1/L$. In this case, the Rabi splitting remains unchanged because the molecular concentration is held fixed. Notably, the driving amplitudes $\eta_j \propto \sqrt{\kappa (L)}$ also depend on $L$. As the cavity length increases, the cavity bandwidth therefore decreases and the drive becomes effectively weaker, which likewise leads to a smaller molecular population and hence a reduced bleach signal.

\end{document}